\documentclass[11pt,letterpaper]{article}
\pdfoutput=1
\usepackage{jheppub}
\usepackage{color}
\usepackage{graphicx}
\usepackage{hyperref}
\usepackage{epsfig}
\usepackage{amsmath}
\usepackage{amssymb}
\usepackage{color}

\newcommand{\mb}[1]{\boldsymbol{#1}}

\newcommand{\be}{\begin{equation}}
\newcommand{\ee}{\end{equation}}

\DeclareRobustCommand{\Sec}[1]{Sec.~\ref{#1}}

\DeclareRobustCommand{\App}[1]{App.~\ref{#1}}

\DeclareRobustCommand{\Fig}[1]{Fig.~\ref{#1}}

\DeclareRobustCommand{\Eq}[1]{Eq.~(\ref{#1})}
\DeclareRobustCommand{\Eqs}[2]{Eqs.~(\ref{#1}) and (\ref{#2})}
\DeclareRobustCommand{\Eqss}[3]{Eqs.~(\ref{#1}), (\ref{#2}), and (\ref{#3})}
\DeclareRobustCommand{\Ref}[1]{Ref.~\cite{#1}}
\DeclareRobustCommand{\Refs}[1]{Refs.~\cite{#1}}

\newcommand{\thetabar}{\overline{\theta}}

\newcommand{\xibar}{\overline{\xi}}

\newcommand{\sigmabar}{\overline{\sigma}}

\newcommand{\psibar}{\overline{\psi}}
\newcommand{\phidot}{\dot{\varphi}}
\newcommand{\vevphi}{\langle \phidot \rangle}

\newcommand{\kahler}{K\"{a}hler }

\newcommand{\BNL}{\mb{B}_{NL}}
\newcommand{\XNL}{\mb{X}_{NL}}

\newcommand{\ANL}{\mb{\mathcal{A}}_{NL}}
\newcommand{\Mpl}{M_{\rm pl}}

\def\ie{{\it i.e.}}
\definecolor{darkgreen}{rgb}{0,.5,0}

\begin{document}

\title{The goldstone and goldstino of supersymmetric inflation}
\author[]{Yonatan Kahn,}
\author[]{Daniel A. Roberts,}
\author[]{and Jesse Thaler}

\affiliation[]{Center for Theoretical Physics, Massachusetts
  Institute of Technology,\\Cambridge, MA 02139, USA}

\emailAdd{ykahn@mit.edu}
\emailAdd{drob@mit.edu}
\emailAdd{jthaler@mit.edu}

\date{\today}

\abstract{We construct the minimal effective field theory (EFT) of supersymmetric inflation, whose field content is a real scalar, the goldstone for time-translation breaking, and a Weyl fermion, the goldstino for supersymmetry (SUSY) breaking.   The inflating background can be viewed as a single SUSY-breaking sector, and the degrees of freedom can be efficiently parameterized using constrained superfields. Our EFT is comprised of a chiral superfield $\XNL$ containing the goldstino and satisfying $\XNL^2 = 0$, and a real superfield $\BNL$ containing both the goldstino and the goldstone, satisfying $\XNL \BNL = \BNL^3 = 0$.  We match results from our EFT formalism to existing results for SUSY broken by a fluid background, showing that the goldstino propagates with subluminal velocities.  The same effect can also be derived from the unitary gauge gravitino action after embedding our EFT in supergravity.  If the gravitino mass is comparable to the Hubble scale during inflation, we identify a new parameter in the EFT related to a time-dependent phase of the gravitino mass parameter.  We briefly comment on the leading contributions of goldstino loops to inflationary observables. 
}

\arxivnumber{}

\preprint{MIT-CTP 4660}

\maketitle

\section{Introduction}

Inflation has emerged as one of the most promising paradigms for the evolution of the early universe  \cite{Guth:1980zm,Linde:1981mu}. Similarly, supersymmetry (SUSY) is a leading candidate for solving many theoretical issues surrounding the Standard Model, including the hierarchy problem and the possible unification of couplings.  Much work has gone into developing concrete models of inflation \cite{Lyth:1998xn} and their embedding into supergravity \cite{2011CQGra..28j3001Y} and string theory \cite{Kachru:2003aw,Balasubramanian:2005zx,Baumann:2014nda} (which necessarily requires SUSY). Each model gives definite predictions for physical observables, but absent smoking-gun evidence for any particular model, it is worth investigating the most general consequences of the simultaneous presence of inflation and SUSY in the early universe. This can be accomplished with an effective field theory (EFT).  EFTs for inflation were originally developed in the non-SUSY case in \Ref{Cheung:2007st} and in the SUSY case in \Refs{Senatore:2010wk,Baumann:2011nk,Baumann:2011nm}.

A convenient way to identify the appropriate low-energy degrees of freedom of an EFT is through an analysis of broken symmetries.    In the presence of SUSY, inflation spontaneously breaks two symmetries of nature. First, inflation picks out a preferred ``direction'' in time (set by the nonzero value of $\frac{1}{2}\dot{\varphi}^2$, the inflaton kinetic energy), spontaneously breaking the diffeomorphisms corresponding to time translations.  Second, the positive vacuum energy of the quasi-de Sitter background necessarily breaks SUSY.  Goldstone's theorem and its SUSY generalization thus imply the existence of two massless modes, the goldstone $\pi$ for time translation breaking and the goldstino $\xi$ for SUSY breaking. These two modes are tied together in an interesting and nontrivial way due to the fact that the spontaneous breaking of Lorentz symmetry (inherited from the breaking of time diffeomorphisms) also breaks SUSY. The fields $\pi$ and $\xi$ are the minimal degrees of freedom necessary for a SUSY EFT of inflation.\footnote{In the context of supergravity (SUGRA), the goldstino is eaten by the gravitino to become its longitudinal mode. The goldstino equivalence theorem implies that at energies $E \gg m_{3/2}$, the gravitino couplings will be dominated by the longitudinal mode, which do not suffer from Planck suppression.} They capture the leading low-energy dynamics of any UV-complete model of SUSY single-field slow-roll inflation, describing fluctuations about a fixed Friedmann-Robertson-Walker (FRW) background.

In this paper, we show that the goldstone and the goldstino, and no other fields, are sufficient to parameterize the minimal degrees of freedom for a SUSY version of the EFT of inflation. Throughout this paper we focus on slow-roll inflation for simplicity, though in principle the formalism can be adapted for more general inflationary scenarios.  Since SUSY requires chiral multiplets to be organized in terms of complex scalars rather than real scalars, an extra scalar partner of the inflaton seems to be a generic consequence of SUSY (as well as a fermionic partner of the inflaton if the inflaton multiplet is separate from the SUSY-breaking multiplet).  For example, \Ref{Baumann:2011nk} constructs a (non-minimal) SUSY EFT of inflation and studies the interplay between the inflaton and the extra scalar.  Here, however, we demonstrate how to consistently decouple the expected extra states, leading to a parametrization of the low-energy degrees of freedom using the constrained supermultiplet formalism of \Ref{Komargodski:2009rz}.\footnote{Nonlinear multiplets were also introduced in \Ref{Cheung:2010mc} in the context of multiple goldstini. For other work involving constrained superfields during inflation, see \Refs{AlvarezGaume:2010rt,AlvarezGaume:2011xv,AlvarezGaume:2011db,Ellis:2013zsa,Antoniadis:2014oya,Ferrara:2014kva,Kallosh:2014via,DallAgata:2014oka,Kallosh:2014hxa,Mavromatos:2014yaa}.} This gives a consistent EFT of $\pi$ and $\xi$ alone.   

Since the dynamics of the goldstone mode is well-studied in the inflation literature, we will take a particular interest in the dynamics of the goldstino.  Recall that the scale of inflation is set by the Hubble parameter $H$.  The scale $H$ is unknown, although it is less than the Planck scale, $H<\Mpl$, possibly even much less.  Thus, one should be skeptical that the effects of the goldstino $\xi$ could be observable, since fermions only contribute to inflationary observables at loop level. Indeed, we estimate that the effect of goldstino loops on the inflaton 2-point function are suppressed by $H^2/\Mpl^2$, and therefore extremely small.  There are potentially observable non-Gaussian signatures in the inflaton 3-point function from goldstino loops, including non-analytic power law dependences and oscillatory behavior, but these are slow-roll suppressed and possibly exponentially suppressed for large $m_{3/2}$ \cite{Arkani-Hamed:2015bza}.  Very optimistically, if the scale of inflation were large (as was suggested by the recent BICEP-2 $B$-mode detection \cite{Ade:2014xna}, though an inflationary interpretation has now been ruled out \cite{Ade:2015tva}), goldstino loop corrections could be measured alongside parametrically similar contributions from graviton loops \cite{Senatore:2009cf}, and the behavior of the 3-point function could confirm the goldstino nature of the exchanged fermions.  This could provide a model-independent diagnostic of the presence of SUSY in our universe, independently of the scale of SUSY breaking today.  SUSY could thus be observable cosmologically even if superpartners are too heavy to be produced at terrestrial colliders.  This point of view has been emphasized in \Refs{Chen:2009zp,2012JCAP...08..033S,2012JCAP...08..019N,Assassi:2013gxa,Craig:2014rta, Arkani-Hamed:2015bza,Dimastrogiovanni:2015pla,Kehagias:2015jha}, which focus on distinctive signatures of Hubble-scale particles during inflation.

The main result of our formalism is the leading-order Lagrangian describing interactions of the goldstone and goldstino in minimal slow-roll inflation, whose \kahler potential $\mb{K}$ and superpotential $\mb{W}$ are given by
\be
\mb{K} = \XNL^\dagger \XNL + \BNL^2, \qquad \mb{W} = \Mpl^2 e^{-i\alpha \ANL/2\Mpl^2 }|m_{3/2}| + f\XNL.
\label{eq:Kcanonicalintro}
\ee
Here and throughout, boldface symbols will indicate superfields. $\XNL$ is the standard constrained chiral supermultiplet from \Ref{Komargodski:2009rz} containing only the goldstino. $\ANL$ is a constrained chiral superfield containing the goldstino and goldstone, and $\BNL$ is a real superfield defined by $\BNL = \frac{1}{2i}(\ANL - \ANL^\dagger)$. $\XNL$ and $\BNL$ form nonlinear representations of SUSY through the constraints
\be
\label{eq:constraintsintro}
\XNL^2 = \XNL \BNL = \BNL^3 = 0.
\ee
Evaluated on an inflating background, the contribution of $\XNL$ to the vacuum energy is positive, while the contribution of $\BNL$ is negative. Both supermultiplets are necessary for a consistent EFT and their individual contributions to the vacuum energy correspond to the inflaton potential and kinetic energy, respectively. The vacuum expectation value (vev) of $\BNL^2$ is proportional to $\dot{H}$ and parameterizes the breaking of time-translation invariance; since no UV completion of spontaneous time translation breaking is known, $\langle \BNL \rangle = \theta \sigma^0 \thetabar \vevphi$ must be imposed by hand.\footnote{Other theories of spontaneous time translation breaking include the (gauged) ghost condensate \cite{ArkaniHamed:2003uy,ArkaniHamed:2004ar,ArkaniHamed:2005gu,Cheng:2006us} (see also \cite{Khoury:2010gb,Koehn:2012te} for extensions to supersymmetry and SUGRA) and time crystals \cite{Shapere:2012nq,Wilczek:2012jt} (see also \cite{PhysRevLett.111.070402}).  These theories describe fluctuations around a fixed vacuum state that spontaneously breaks time translations, whereas the EFT of inflation describes fluctuations around the classical solution for the slowly-rolling inflaton.  Despite the mismatch between a fixed vacuum and a slowly-rolling ``vacuum'', the inflaton fluctuations can still be described by the goldstone $\pi$ from spontaneous symmetry breaking.}

The Lagrangian in \Eq{eq:Kcanonicalintro} contains three free parameters beyond $\vevphi$,
\be
\{ f, |m_{3/2}|, \alpha \}.
\label{eq:params}
\ee
$f$ fixes the positive contribution to the vacuum energy from the inflaton potential, and $m_{3/2} = e^{-i \alpha \vevphi t / 2 \Mpl^2} |m_{3/2}|$ is the mass of the gravitino.\footnote{We will be doggedly insistent on writing $|m_{3/2}|$, rather than $m_{3/2}$, for the magnitude of the gravitino mass. The reason is that time diffeomorphism breaking makes possible a nontrivial time-dependent phase on $m_{3/2}$.}  In the case where $|m_{3/2}| \ll H$, the minimal EFT of inflation is parameterized only by $f$ and $\vevphi$, and the leading effects of the theory can be captured in the framework of rigid SUSY. However, in the case where the gravitino mass is large compared to the other scales in the EFT, the theory must be coupled to SUGRA. In that case, in addition to the new scale $|m_{3/2}|$, we find an additional free parameter $\alpha$. To our knowledge, $\alpha$ represents a previously unknown inflationary parameter, absent in the rigid SUSY limit.\footnote{As discussed in \Sec{sec:linearframe}, $\alpha$ can related to a non-zero vev for the vector auxiliary field $b_\mu$ of SUGRA.}  Through an analysis of the goldstone-goldstino couplings, we argue that the natural scale to suppress additional higher-dimension operators in the EFT is
\be
\label{eq:naturalcutoff}
\Lambda = \sqrt{\Mpl H}.
\ee

\begin{figure}[t]
\begin{center}
\includegraphics[width=0.9\columnwidth]{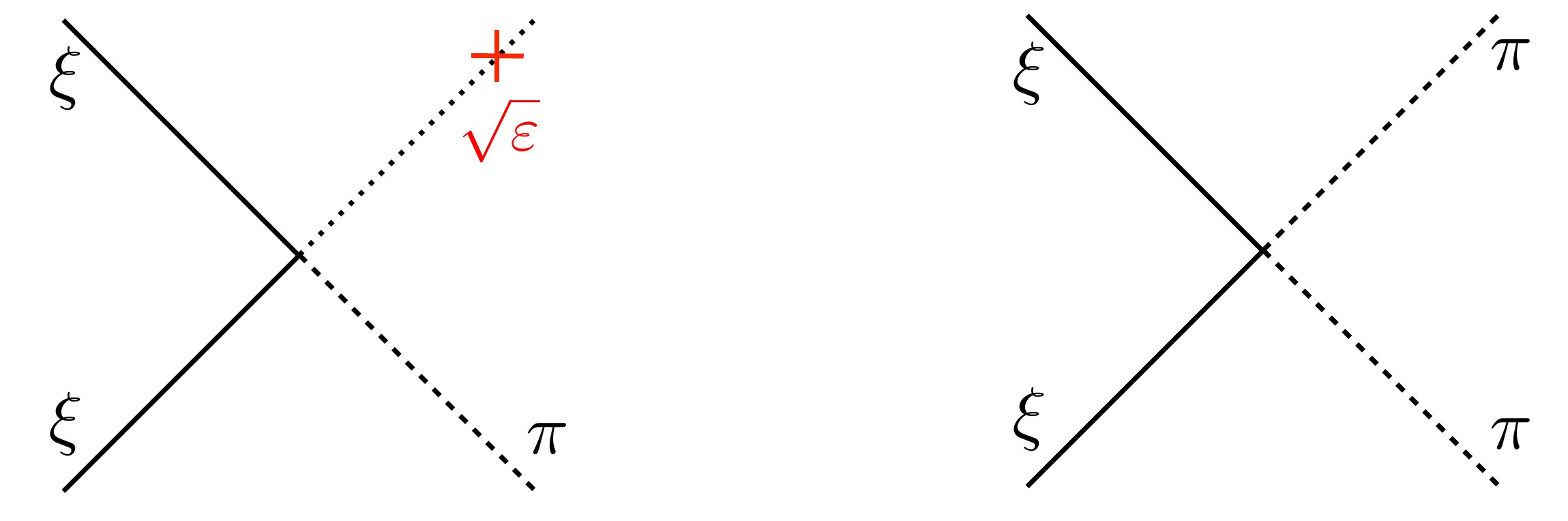}
\caption{Schematic Feynman diagrams for the leading goldstone-goldstino interactions. 3-point vertices are suppressed by the slow-roll parameter $\sqrt{\varepsilon}$ compared to 4-point vertices when the inflaton leg is put to its vev. It is  analogous to the way vertices in the Standard Model with one Higgs boson have an additional factor of the Higgs vev $v$ compared to vertices with two Higgs bosons. Since fermions are unobservable during inflation, these interactions contribute to correlators of $\pi$ through loops of $\xi$.}
\label{fig:3and4pt}
\end{center}
\end{figure}

This minimal EFT of inflation contains a Lorentz-violating kinetic term for the goldstino, and derivative interactions between $\xi$ and $\pi$, shown schematically in \Fig{fig:3and4pt}. The goldstino has a relativistic dispersion relation $\omega = c_s k$, with a nontrivial speed of sound
\be
c_s = 1 - \frac{2}{3}\varepsilon,
\ee
where $\varepsilon$ is the slow-roll parameter
\be
\varepsilon \equiv - \frac{\dot{H}}{H^2} = \frac{\vevphi^2}{2 \Mpl^2 H^2}. \label{eq:slowroll}
\ee
This reduced speed of sound is known in the literature as the ``slow gravitino'' \cite{Benakli:2013ava, Benakli:2014bpa}, which we relate to inflationary dynamics and which can be derived directly from our formalism.  

We use the goldstone-goldstino interactions derived from \Eq{eq:Kcanonicalintro} to estimate the loop contributions to the two-point function $\langle \pi \pi \rangle$, which is related to the correlation function of primordial curvature perturbations. These loop effects are suppressed compared to the tree-level result by a factor of 
\be
\frac{\delta \langle \pi \pi \rangle}{\langle \pi \pi \rangle_0 } \sim \frac{H^4}{\Lambda^4} = \frac{H^2}{\Mpl^2} ,
\ee
similar to expectations from non-SUSY gravitational contributions \cite{Senatore:2009cf}. The contributions from $\xi$ are thus extremely small unless $H$ is large.  As noted in \Ref{Arkani-Hamed:2015bza}, the mass and spin of $\xi$ may give distinctive signatures to the 3-point function $\langle \pi \pi \pi \rangle$, and in particular the non-standard dispersion relation for $\xi$ may signal its goldstino nature and stand out from the effects of loops of other fermions.

The rest of this paper is organized as follows. In \Sec{sec:Review}, we briefly review the standard EFT of single-field inflation. In \Sec{sec:EFT}, we build the minimal EFT of SUSY inflation in the rigid flat-space limit, first constructing the nonlinear fields $\XNL$ and $\BNL$, and then constructing the lowest-order Lagrangian, which describes minimal slow-roll inflation. We also show how the constraints on $\BNL$ can be derived by integrating out the extra states at the cutoff scale $\Lambda$ from a generic chiral multiplet containing the inflaton. In \Sec{sec:kinetic}, we investigate the dynamics of the goldstino in rigid flat space, including the goldstino dispersion relation and the leading goldstone-goldstino interactions. As a nontrivial check of our formalism, we show that the goldstino kinetic term matches existing results on SUSY broken by a fluid background. In \Sec{sec:SUGRA}, we relax from the rigid SUSY limit and write down our EFT in SUGRA. We estimate the leading effects of the new parameter $\alpha$, and following \Ref{Giudice:1999yt}, we work out the gravitino mode equations in an FRW background and match to the goldstino results.  Finally, we estimate the goldstino contribution to loop corrections of inflationary observables.  We conclude in \Sec{sec:conclusion}. Some technical details are left to the appendices. We follow the conventions of \Ref{wess1992supersymmetry} throughout, with the exception of the Ricci scalar, where we pick the opposite sign.

\section{Review of EFT of inflation}
\label{sec:Review}

Here we present a brief summary of the EFT of inflation. Readers already familiar with this subject may wish to skip to \Sec{sec:EFT}. Readers unfamiliar with this subject are encouraged to consult \Ref{Cheung:2007st} for more details.

The idea is to construct an EFT of fluctuations about a fixed FRW background, parameterized by $H$ and $\dot{H}$, sourced by an inflaton $\varphi$. Since we are interested in evaluating inflationary observables at horizon crossing, $\omega \sim H$, the IR cutoff of the theory should be $H$. The time dependence of the background, parameterized by $\dot{H}$, spontaneously breaks time diffeomorphisms $t \to t + \lambda(t, \vec{x})$, while the spatial homogeneity of the background preserves time-dependent spatial diffeomorphisms. The action for such an EFT can contain terms invariant under time-dependent spatial diffeomorphisms but not time diffeomorphisms, for example $g^{00}$. It was shown in \Ref{Cheung:2007st} that the most general action is a polynomial in two objects, $g^{00} + 1$ and $\delta K_{\mu \nu}$, where the latter is the perturbation of the extrinsic curvature of surfaces of constant time compared to the extrinsic curvature of the background. All coefficients quadratic and higher in these objects are free parameters of the EFT, but the constant and linear terms are fixed by the Friedmann equations:
\be
\label{eq:EFTunitary}
S = \Mpl^2 \int d^4x \, \sqrt{-g} \left [ \frac{1}{2} R  - (3H^2 + \dot{H}) +  \dot{H}g^{00} + \cdots \right].
\ee
In this unitary-gauge action, the propagating degrees of freedom are those of the metric, which has acquired a longitudinal mode by eating the scalar perturbation $\delta \varphi$, in analogy to the ordinary Higgs mechanism for spontaneously-broken gauge theories. Said another way, we have chosen a gauge where at all times the inflaton takes its homogeneous background value, so all variations about the background are described by $g^{\mu \nu}$.

The longitudinal mode $\pi$ of the metric can be isolated by performing a broken time diffeomorphism on the unitary gauge action. For simplicity, we only deal with slow-roll inflation in this discussion. The goldstone equivalence theorem implies that at sufficiently high energies, the interactions of a massive gauge field are dominated by its longitudinal mode. In the inflationary case, ``sufficiently high energies'' means at scales above $\omega \sim \sqrt{\varepsilon} H$, where $\varepsilon$ is the slow-roll parameter in \Eq{eq:slowroll}. Since our EFT has an IR cutoff at $\omega \sim H$, we are parametrically well within the goldstone equivalence regime during slow-roll inflation with $\varepsilon \ll 1$,  and the interactions of the metric are dominated by $\pi$; mixing terms such as $\dot{\pi} \delta g^{00}$ can be dropped.\footnote{In more general models of inflation, the mixing with gravity can become relevant at a higher scale and such terms may not be neglected, see e.g.~\cite{Cheung:2007st,Senatore:2010wk,Senatore:2009gt}.} In an inflationary context, this is known as the \emph{decoupling limit}. Performing the broken time diffeomorphism on \Eq{eq:EFTunitary} gives
\be
S = \Mpl^2 \int d^4x \, \sqrt{-g} \left [ \frac{1}{2} R  - (3H^2 + \dot{H}) - \dot{H} - \frac{1}{2}\partial_\mu \pi \partial^\mu \pi + \cdots \right].
\label{eq:EFTpi}
\ee
The first term is the gravity action, and the second and third terms correspond to the inflaton potential and kinetic energy, respectively. Crucially, the term $\dot{H}g^{00}$ in \Eq{eq:EFTunitary} also generates a kinetic term for $\pi$, which is evaluated with respect to the background FRW metric. Additional interactions for $\pi$ are encoded in higher-dimensional operators in the unitary gauge action. 

As an example of this formalism applied to slow-roll inflation \cite{Cheung:2007st, Baumann:2011nk}, consider the Lagrangian for the inflaton $\varphi$:
\be
\label{eq:EFTphi}
\mathcal{L} = -\frac{1}{2}g^{\mu \nu}\partial_\mu \varphi \partial_\nu \varphi -V(\varphi).
\ee
Since the gauge choice required for unitary gauge is a spacetime-dependent shift in $t$, we can apply the broken time diffeomorphism directly to \Eq{eq:EFTphi} by
\be
\varphi \to \vevphi t + \pi(x).
\label{eq:timediff}
\ee
Unitary gauge here corresponds to $\pi(x) = 0$ identically, where $\varphi$ follows its background solution at all times and all fluctuations are in the metric. The Friedmann equations fix
\begin{align}
V(\varphi) & = \Mpl^2(3H(\varphi)^2 + \dot{H}(\varphi)), \label{eq:Vphi} \\
\vevphi & = \Mpl \sqrt{ 2 |\dot{H}(\varphi)|},
\label{eq:phidotHdot}
\end{align}
for a given choice of FRW background parameterized by $H$ and $\dot{H}$. In the decoupling limit, we can evaluate the $\varphi$ kinetic term on the background metric. Under the further assumption that any time variation in $H$ or $\dot{H}$ is slow-roll suppressed, we can replace $H^2(\varphi)$ and $\dot{H}(\varphi)$ by their background values $H^2$ and $\dot{H}$. This brings the action to the form of \Eq{eq:EFTpi}.\footnote{A term $\dot{H} \dot{\pi}$ can be dropped because it is a total time derivative, under the assumption that $\dot{H}$ is constant.}

\section{Minimal EFT of supersymmetric inflation}
\label{sec:EFT}

Generalizing the approach in the previous section, we wish to construct a SUSY EFT valid at scales $H \lesssim \omega < \Lambda$ whose low-energy degrees of freedom are a single real scalar and a Weyl fermion: the inflaton fluctuations $\pi$ and the goldstino $\xi$.  As anticipated in \Eq{eq:naturalcutoff}, we find the natural cutoff scale is $\Lambda = \sqrt{H \Mpl}$.  Clearly there is a mismatch between bosonic and fermionic degrees of freedom, but this is not an obstacle to building a SUSY theory; the constrained superfield formalism of \Ref{Komargodski:2009rz} allows the construction of composite superfields using the goldstino and its bilinears as components, which evades the problem. Our formalism will demonstrate that two logically distinct sources of SUSY breaking, namely Lorentz breaking and vacuum energy, can be viewed as coming from a single SUSY-breaking sector, the inflaton background. In this and the following section, we consider the theory in the rigid flat-space limit, but in \Sec{sec:SUGRA} we will couple the theory to SUGRA in an FRW background.

\subsection{Constrained superfields}
\label{sec:constrained}

As noted in \Ref{Baumann:2011nk}, the positive vacuum energy of the inflaton potential spontaneously breaks SUSY, and thus there must be a goldstino. We parameterize it by a chiral superfield
\be
\XNL = \frac{\xi^2}{2F} + \sqrt{2}\theta \xi + \theta^2 F
\label{eq:xnl-def}
\ee
satisfying
\be
\XNL^2 = 0.
\label{eq:XConstraint}
\ee
Here $F$ is not fixed but is an auxiliary field whose vev we will want to set dynamically to $\langle F \rangle = -f$. We take $F$ to be real, since we can absorb any relative phase in $\langle F \rangle$ with a field redefinition, as we will demonstrate in \Eq{eq:Xredef} below. 

To parametrize the inflaton, we start with the constrained ``axion'' superfield $\ANL$ of \Ref{Komargodski:2009rz}, which describes a single real scalar $\varphi$ with a shift symmetry $\varphi \to \varphi + c$.
Its lowest component is
\be
\ANL | = \varphi + \frac{i}{2F^2}\xi \sigma^\mu \xibar \partial_\mu \varphi;
\ee
note that it contains both the inflaton $\varphi$ and the goldstino $\xi$. The constraint equation for $\ANL$ is
\be
\XNL(\ANL - \ANL^\dagger) = 0
\label{eq:XAConstraint}
\ee
which also implies
\be
(\ANL - \ANL^\dagger)^3 = 0.
\ee
In what follows, we demand that our EFT exhibits manifest shift symmetry for $\varphi$ (and hence for the goldstone $\pi$), and so we will express them as a function of the real superfield
\be
\BNL \equiv \frac{1}{2i}(\ANL - \ANL^\dagger),
\label{eq:BNL-def}
\ee
satisfying
\be
\label{eq:BConstraints}
\BNL^3 = 0, \qquad \XNL \BNL = 0.
\ee
The expression of $\BNL$ in terms of its component fields is given in \App{app:BNL}. The precise form is not particularly illuminating, so we refrain from writing it out in full here and refer to its components when necessary. The component of $\BNL$ containing only the inflaton is
\be
\BNL |_\varphi = \theta \sigma^\mu \thetabar \, \partial_\mu \varphi.
\ee
When we introduce broken time diffeomorphisms by
\be
\partial_\mu \varphi \to \vevphi \delta_\mu^0 + \partial_\mu \pi,
\label{eq:brokentimediff}
\ee
with $\vevphi = \Mpl \sqrt{ 2 |\dot{H}|}$ as before, a term $\BNL^2$ in the \kahler potential will generate a canonically normalized kinetic term for $\pi$.

Implicit in the above discussion is that we are working in the slow-roll limit where the constraints in \Eqs{eq:XConstraint}{eq:BConstraints} can be regarded as being independent of the space-time background.  To treat more general inflationary scenarios, we would need to account for the possibility that these constraints depend on the gravity multiplet as well.  We leave such generalizations to future work.

The reader may wonder why, in the spirit of minimality, we need the extra multiplet $\XNL$ in the Lagrangian in addition to $\BNL$, which already contains both of our low-energy fields. A Lagrangian built solely out of $\BNL$ is actually pathological, for two related reasons. First, when we impose a vev for $\phidot$ to obtain an inflationary background, the \kahler potential $\BNL^2$ which gives the $\pi$ kinetic term also contributes a vacuum energy
\be
\mathcal{L} \supset \int d^4 \theta \, \langle \BNL \rangle ^2 =  -\frac{1}{2}\langle \partial_\mu \varphi \rangle \langle \partial^\mu \varphi \rangle = +\frac{1}{2}\vevphi^2.
\label{eq:LorentzVacEnergy}
\ee
This positive contribution to the Lagrangian is a negative contribution to the Hamiltonian, and thus the vacuum energy from Lorentz breaking is negative. This is already pathological at the level of the flat-space SUSY algebra, which requires all states to have non-negative energy. Second, one can see from the component expansion of $\BNL$ that any kinetic structure for the goldstino will be proportional to $\partial_\mu \varphi \partial^\mu \varphi$, and hence will vanish in the pure de Sitter limit of $\langle \phidot \rangle = 0$.\footnote{Note that in general inflationary scenarios beyond slow-roll, $\vevphi = 0$ does not correspond to a pure de Sitter limit \cite{Cheung:2007st}.}$^,$\footnote{One can attempt to dispense with the constraint $\XNL \BNL = 0$ and solve only the constraint $\BNL^3 = 0$ directly. Unlike the $\BNL$ above, this gives a pure fermion $\theta \xi$ in the $\theta$ component. However, this theory is plagued by its own pathologies, including kinetic terms which are necessarily nonlocal in superspace, and superluminal goldstino propagation.}

\subsection{Lowest-order Lagrangian}
\label{sec:lowestSUSY}

Thanks to the constraints in \Eq{eq:BConstraints}, the possible Lagrangians we can write down are extremely restricted. To lowest order in derivatives, the most general Lagrangian in terms of $\XNL$ and $\BNL$ is
\be
\label{eq:generalLag}
\mathcal{L} =  f \int d^2 \theta \, \XNL + {\rm h.c.} + \int d^4 \theta \, \left \{ \XNL^\dagger \XNL + \BNL^2 + \alpha \BNL + \beta \XNL + \beta^* \XNL^\dagger \right\}.
\ee
The dimension-1 coefficients $\alpha$ and $\beta$ are free parameters; we include them even though they multiply linear terms which integrate to zero in flat space because they can contribute after coupling to SUGRA.\footnote{Note in particular that we cannot get rid the linear term in $\BNL$ by completing the square and shifting $\BNL$ by a constant, because that would be inconsistent with the constraint $\BNL^3 = 0$.} As we will see in \Sec{sec:SUGRA}, $\beta$ can be removed by a \kahler transformation in SUGRA, but $\alpha$ gives qualitatively new effects which are absent in rigid SUSY. 

The coefficient of $\XNL^\dagger \XNL$ is fixed by requiring canonical normalization for the goldstino when $\vevphi = 0$. By \Eq{eq:LorentzVacEnergy}, in this limit the Lagrangian (\ref{eq:generalLag}) is simply the Polonyi model, where the goldstino must be canonically normalized. The breaking of time diffeomorphisms offers the possibility of a time-dependent phase of the term $f \XNL$ in the superpotential, of the form
\be
\mathcal{L} \supset \int d^2 \theta \,  f e^{i\gamma\ANL/\Mpl^2} \XNL.
\ee
However, this phase can be absorbed into a field-dependent redefinition of $\XNL$,
\be
\XNL \to e^{-i\gamma\ANL/\Mpl^2} \XNL,
\label{eq:Xredef}
\ee
which is compatible with the constraints of \Eqs{eq:XConstraint}{eq:XAConstraint} and leaves the \kahler potential $\XNL^\dagger \XNL$ invariant. Thus the phase $\gamma$ is unphysical, and without loss of generality, we can take $f$ to be real. On the other hand, we will see in \Sec{sec:kahler} that $\alpha$ gives rise to an analogous phase on the constant term in the SUGRA superpotential, which cannot be removed.

The remaining coefficients in \Eq{eq:generalLag} are fixed to ensure that we get the correct inflaton Lagrangian during inflation. A canonically normalized inflaton kinetic term fixes the coefficient of $\BNL^2$ to be unity from \Eq{eq:LorentzVacEnergy}. Finally, we identify
\be
f = \sqrt{V(\varphi)}, \label{f-indentification}
\ee
where it is understood that $\varphi$ is evaluated on its background slow-roll solution.
  In terms of the FRW parameters, we can also write
\be
f = \Mpl H\sqrt{3 - \varepsilon},
\label{f-id-H}
\ee
which comes from \Eq{eq:Vphi}.  Note that $f$ is proportional to $\Lambda^2 = H \Mpl$, as expected since $f$ is the goldstino decay constant.  The relation (\ref{f-indentification}) always holds in SUGRA for arbitrary $m_{3/2}$, but \Eq{f-id-H} is only true in SUGRA when $m_{3/2}=0$.  When we discuss goldstino dynamics in \Sec{sec:kinetic}, we consider this theory in the rigid flat space limit, where $\Mpl \to \infty$ and the nonzero background value of the inflaton stress-energy tensor does not back-react on the geometry.

\subsection{Higher-order effects}

It is simple to extend the lowest-order Lagrangian,  \Eq{eq:generalLag}, to include higher-order effects in the EFT by adding terms with superspace or spacetime derivatives.  Such terms are expected to be suppressed by powers of $\Lambda$ and can be important for inflationary phenomenology, as emphasized in \Refs{Baumann:2011nk,Baumann:2011nm,Khoury:2010gb,Sasaki:2012ka,Gwyn:2014wna}.  For example, a nontrivial speed of sound $c_\pi$ for the goldstone can be obtained by adding self-interactions. A good candidate higher-derivative operator is $\BNL^2 \partial_\mu \ANL \partial^\mu \ANL^\dagger$, whose top component contains $(\partial_\mu \varphi \, \partial^\mu \varphi )^2$. However, such an operator contributes to the vacuum energy, so adding it by itself will shift the vacuum energy---or, in field theoretic terms, will reintroduce a tadpole for $\pi$. To avoid this, we may only add higher-order terms to the action in the combination
\be
\Big(1 + \frac{1}{ \vevphi^2}\partial_\mu \varphi \partial^\mu \varphi \Big)^k  = \mathcal{O}(\pi^2),
\ee 
 for some power $k>1$, since that combination is quadratic in the fluctuations. 

 In the original EFT of inflation, to get a nontrivial speed of sound for $\pi$, one adds \cite{Cheung:2007st} 
\be
\Delta \mathcal{L} = M^4 \Big( 1 + \frac{1}{ \vevphi^2}\partial_\mu \varphi \partial^\mu \varphi \Big)^2, \label{eq:higher-derivative}
 \ee
where the speed of sound is given in terms of the new mass parameter $M$ and the parameters of the background as
\be
c_\pi = \bigg(1 + \frac{4M^4}{\Mpl^2 |\dot{H}|} \bigg)^{-1/2}. \label{eq:goldstone-speed-of-sound}
\ee
Such an operator, however, does not exist by itself in a SUSY theory \cite{Baumann:2011nk}. Expanding out \Eq{eq:higher-derivative} and comparing with \Eq{eq:generalLag}, we see that the operator we must add to shift the speed of sound as \eqref{eq:goldstone-speed-of-sound} while preserving the background is
\begin{align}
\Delta \mathcal{L}  = & -\frac{M^4}{2\Mpl H\sqrt{3-\varepsilon}} \int d^2 \theta \, \XNL + {\rm h.c.} \nonumber \\ 
& - \frac{M^4}{\Mpl^2 |\dot{H}|}\int d^4 \theta \, \left \{ \BNL^2 \Big(2+ \frac{1}{2\Mpl^2 |\dot{H}| } \partial_\mu\ANL \partial^\mu \ANL^\dagger\Big) \right \}.
\end{align}
This is very similar to the operator used in \Ref{Baumann:2011nk} to construct a SUSY theory with small sound speed, with the constrained superfield $\ANL$ here playing the role of the chiral multiplet where the inflaton resides. However, since the inflaton multiplet of \Ref{Baumann:2011nk} is not constrained, the structure of the operator here is a little different.   Note that these operators are indeed suppressed by powers of $\Lambda$.

\subsection{Connection to previous literature}
We can make an even more direct connection between the theory we study here and that of \Ref{Baumann:2011nk} by showing how to derive the $\ANL$ constraints. In particular, we can think of our constraints as arising from integrating out the additional real scalar and Weyl fermion of a chiral inflaton multiplet. Consider two chiral multiplets, $\mb{A}$ and $\XNL$, as in \Ref{Baumann:2011nk}. $\XNL$ is a constrained superfield of $F$-term breaking, containing only the goldstino and satisfying $\XNL^2=0$, and $\mb{A}$ is a priori unconstrained. $\mb{A}$ contains the inflaton $\varphi$ and a scalar partner $\sigma$ in its lowest component, as well as an additional fermion $\psi$:
\be
\mb{A} =\varphi + i\sigma + \sqrt{2}\theta \psi + \theta^2 F_A.
\ee 

Much of the focus of \Ref{Baumann:2011nk} was on the effects of $\sigma$, though such a mode is not necessary in our minimal EFT. From a Planck-suppressed coupling of the form
\be
\label{eq:decoup}
\frac{\lambda_\sigma}{24\Mpl^2} \int d^4 \theta \, (\mb{A} - \mb{A}^{\dagger})^2 \XNL^{\dagger} \XNL \supset -\frac{1}{2} \lambda_\sigma H^2\left(1 - \frac{\varepsilon}{3}\right) \sigma^2,
\ee
$\sigma$ receives a mass $m_\sigma^2 =  \lambda_\sigma H^2$ (plus slow-roll corrections), where $\lambda_\sigma$ is a dimensionless constant. For $\lambda_\sigma$ of order unity, the mass of the extra scalar is at the EFT scale $H$. For $\lambda_\sigma \gg 1$, however, $\sigma$ decouples and we may integrate it out.   In particular, if we rewrite
\be
\frac{\lambda_\sigma}{\Mpl^2} \Rightarrow  \frac{\lambda'_\sigma}{\Lambda^2},
\ee
then for $\lambda'_\sigma$ being order 1, $m_\sigma$ is order $\Lambda = \sqrt{\Mpl H}$ as suggested by \Eq{eq:naturalcutoff}.  Thus, $\Lambda$ is the natural cutoff scale to suppress higher-dimension operators in our SUSY EFT.

In \Ref{Baumann:2011nk} it was also noted that $\psi$ acted as an additional goldstino, interpreting $\mb{A}$ and $\XNL$ as separate sources of SUSY-breaking for Lorentz breaking and vacuum energy, respectively. Here, we can consistently decouple $\psi$ with the higher-dimension operator
\be
-\frac{\lambda_\psi}{12 \Mpl^3}\int d^4 \theta \, \left \{ \overline{D}_{\dot{\alpha}}{\mb A}^\dagger \overline{D}^{\dot{\alpha}}{\mb A}^\dagger + D^\alpha{\mb A} D_\alpha {\mb A} \right \} \XNL^{\dagger} \XNL \supset -\frac{1}{2} \lambda_\psi \frac{H^2}{\Mpl} \left(1 - \frac{\varepsilon}{3}\right) \left ( \psi^2 + \bar{\psi}^2 \right)
\label{eq:LambdaDecoup}
\ee
for sufficiently large $\lambda_\psi$. From the power counting of \Ref{Baumann:2011nk}, this fermion mass is expected to be Planck-suppressed, so a heavy fermion from large $\lambda_\psi$ is not natural in that context.  For the present EFT, though, $\mb{A}$ is not a fundamental degree of freedom in the theory below $\Lambda$.  In particular, making the replacement $\lambda_\psi/\Mpl^3 \Rightarrow \lambda_\psi' / \Lambda^3$, then the additional fermion is again at the cutoff scale $\Lambda$ for $\lambda_\psi'$ being $\mathcal{O}(1)$.

Together, the higher-dimensional operators in \Eqs{eq:decoup}{eq:LambdaDecoup} impose the constraint
\be
\XNL (\mb{A} - \mb{A}^{\dagger}) = 0
\ee
when $\lambda_\sigma, \lambda_\psi \gg 1$.
Solving this constraint for $\mb A$ as in \Ref{Komargodski:2009rz}, we recover our nonlinear superfield $\ANL$ with the constraint in \Eq{eq:XAConstraint}. Note that the requirement of manifest shift symmetry in $\varphi$ protects the mass of $\varphi$, but not the mass of any other components of $\mb{A}$.

\section{Goldstino dynamics in rigid flat space}
\label{sec:kinetic}

The nonzero vev $\vevphi$ spontaneously breaks Lorentz invariance, and we expect this Lorentz breaking to appear in the goldstino kinetic term as well as in the goldstone-goldstino interactions. To derive these effects, we work in the rigid flat-space limit for simplicity, taking $\Mpl\to\infty$ and considering only modes $\omega \gg H$, such that the background is approximately flat. This corresponds to ignoring any Hubble friction terms in the goldstino dispersion relation.

The key result we will derive from our EFT is that the goldstino has a linear dispersion $\omega = c_s k$ with a speed of sound
\be
c_s = \frac{1 - \kappa}{1+ \kappa} = 1 - \frac{2}{3}\varepsilon,
\label{eq:cs}
\ee
where
\be
\kappa = \frac{\langle \phidot \rangle ^2}{2 \langle F \rangle ^2} = \frac{\varepsilon}{3 - \varepsilon}
\label{eq:kappadef}
\ee
is a dimensionless parameter characterizing the relative size of the two sources of SUSY breaking: Lorentz breaking and vacuum energy. The relationship between $\kappa$ and the slow-roll parameter $\varepsilon$ comes from \Eqs{eq:phidotHdot}{f-id-H}.  This same parameter $c_s$ will appear in \Sec{sec:gravdispersion} when we include the effects of the gravitino mass term.

\subsection{Relevant scales and goldstino equivalence} \label{Sec:scales}

We begin with some comments about the relevant scales in our EFT. The theory we described in \Sec{sec:EFT} has a single sector responsible for SUSY breaking, namely the inflaton background. This sector breaks SUSY through both an $F$-term and Lorentz violation. 
However, we know that SUSY is also broken today, corresponding to a positive vacuum energy $F_0^2$ for a SUSY-breaking scale $F_0$. To compensate for the fact that the vacuum energy is vanishingly small today, we would need to add a negative cosmological constant \cite{DEramo:2013mya,Bertolini:2013via}. The tuning of the cosmological constant to obtain flat space gives a well-known relation between the gravitino mass and the scale of SUSY breaking, $m_{3/2} = F_0/\sqrt{3}\Mpl$.  In the following, we assume that $|m_{3/2}|$ does not change during or after inflation and we use $|m_{3/2}|$ as a proxy for $F_0$ when comparing different scales in the EFT. We emphasize, though, that $|m_{3/2}|$ is not an order parameter for SUSY breaking during inflation.

If $F_0$ is much less than the other scales in the EFT,
\be
\label{eq:scales1}
F_0 \ll \langle \phidot \rangle \ll \Lambda^2,
\ee
or equivalently (dividing by $\Mpl$)
\be
\label{eq:scales2}
|m_{3/2}| \ll \sqrt{\varepsilon} H \ll H,
\ee
then we can neglect it. This corresponds to the goldstino equivalence limit, where even loop diagrams in the EFT are dominated exclusively by the goldstino because all momentum scales in the loop are well above the gravitino mass. Indeed, we are already neglecting terms of $\mathcal{O}(\sqrt{\varepsilon})$ mixing $\pi$ with metric fluctuations by working in the decoupling limit, so the effects of $F_0$ are an even smaller perturbation on top of these terms. 

If instead $F_0 \simeq \Lambda^2$, the goldstino equivalence limit is no longer appropriate, and we would need to consider the interactions of the goldstone with the gravitino, as well as goldstino-gravitino mass mixing.\footnote{If $F_0 \simeq \vevphi$, slow-roll suppressed effects are parametrically similar to effects proportional to $m_{3/2}$, and some care would be required in including the leading effects of both.}  Below we will work exclusively in the goldstino equivalence limit given by \Eqs{eq:scales1}{eq:scales2}, leaving a discussion of the case of large $|m_{3/2}|$ for \Sec{sec:SUGRA}.
 
 \subsection{Goldstino dispersion relation from EFT Lagrangian}

The \kahler potential terms in \Eq{eq:generalLag} give kinetic terms for the goldstino after performing the superspace integral. The $\XNL^\dagger \XNL$ term gives the canonical kinetic term for $\xi$, but when $\BNL$ acquires a vev $\langle \BNL \rangle = \theta \sigma^0 \thetabar \vevphi$, the $\BNL^2$ term gives Lorentz-violating corrections to the $\xi$ kinetic term. By inspection, we see that $\BNL^2$ does not contain any goldstino bilinears with no derivatives in its top component, so there is no mass term, as expected for a goldstino. 

The remaining goldstino bilinears are all one-derivative terms arising from expanding the $\theta, \thetabar$ components of $\BNL$, given in \App{app:BNL}:
\be
\label{eq:Bsq}
\BNL^2|_{{\rm kin}}  = \frac{-i \vevphi^2}{2F^2} \left [ \thetabar^2 (\partial_\nu \xi \sigma^0 \sigmabar^\nu \theta)(\theta \sigma^0 \xibar) + \theta^2 (\partial_\nu \xibar \sigmabar^0 \sigma^\nu \thetabar) (\thetabar \sigmabar^0 \xi) \right ].
\ee
Simplifying using the definition of $\kappa$ in \Eq{eq:kappadef}, we find
\be
\mathcal{L} \supset -i\kappa \partial_\mu \xibar \sigmabar^\mu \xi + 2i\kappa \partial_0 \xibar \sigmabar^0 \xi.
\ee
Including the canonical kinetic term originating from $\int d^4 \theta \, \XNL^\dagger \XNL$,
\be
\mathcal{L} \supset i\partial_\mu \xibar \sigmabar^\mu \xi,
\ee
we have the goldstino kinetic structure 
\be
\mathcal{L}_{{\rm kin}} = (1-\kappa)i \partial_\mu \xibar \sigmabar^\mu \xi + 2\kappa i \partial_0 \xibar \sigmabar^0 \xi.
\ee
One can then read off that the goldstino satisfies the dispersion relation $\omega = |c_s| k$ with
\be
c_s = \frac{1 - \kappa}{1+ \kappa}.
\label{eq:cs2}
\ee

Clearly, \Eq{eq:cs2} is pathological when $\kappa \geq 1$, but $\kappa < 1$ is automatically satisfied during inflation. Indeed, since $\varepsilon \ll 1$ during inflation and $\varepsilon = 1$ ends inflation, our EFT of inflation is only valid when $\kappa = \varepsilon/(3 - \varepsilon)$ is much less than $1/2$. However, in a more general context, requiring a goldstino sound speed which does not cross zero can be read as a constraint that the vacuum energy from $F$-term breaking must be larger than the vacuum energy from time diffeomorphism breaking (see \Eq{eq:LorentzVacEnergy}). 
\subsection{Cross-check: Goldstino dispersion relation from a fluid background}

An alternate way to derive \Eq{eq:cs} is to use results from the literature on SUSY breaking at finite temperature \cite{Kratzert:2003cr} and in a fluid background \cite{Benakli:2013ava, Benakli:2014bpa}. In both cases, there is a massless fermionic excitation, the ``phonino'', with a linear dispersion and speed of sound equal to 
\be
c_s = |w| = \left | \frac{p}{\rho} \right |,
\ee
where $p$, $\rho$, and $w$ are the pressure, density, and equation of state of the background fluid, respectively. Both the existence of this mode and its dispersion relation follow from extremely general considerations, namely applying the Ward-Takahashi identity for the supercurrent, which contains the vev of the stress tensor on the right-hand side.

In our case, we simply plug in the stress tensor of the inflaton background:
\be
p_\varphi = \frac{1}{2}\dot{\varphi}^2 - V(\varphi), \qquad \rho_\varphi = \frac{1}{2}\dot{\varphi}^2 + V(\varphi).
\label{eq:prhophi}
\ee
This gives
\be
c_s = \frac{V(\varphi) - 2\phidot^2}{V(\varphi) + 2\phidot^2} = \frac{1 - \kappa}{1 + \kappa}
\ee
after making the identifications in \Eqs{f-indentification}{eq:kappadef}. As expected, this agrees with our EFT derivation.

\subsection{Leading goldstino-goldstone interactions}
\label{sec:interactionsRigid}

Beyond just dispersion relations, our EFT allows us to derive the leading goldstino-goldstone interactions in a model-independent way. We seek terms in $\BNL^2$ containing at most two goldstinos and two $\varphi$ fields. Using the expansion of $\BNL$ in \App{app:BNL}, we find the following:
\be
\label{eq:leadingxiphi}
\mathcal{L} \supset \frac{i}{4F^2}\partial_\rho \varphi \partial_\nu(\xi \partial_\mu \varphi) \sigma^\mu \sigmabar^\nu \sigma^\rho \xibar + {\rm h.c.}
\ee
After the replacement in \Eq{eq:brokentimediff}, the terms where one factor of $\partial \varphi$ is replaced by $\vevphi$ give 3-point interactions, 
\be
\mathcal{L}_{\text{3-pt}} = \frac{\vevphi}{4F^2} \Big(i \, \partial_\rho \pi \, \partial_\nu \xi \, \sigma^0  \sigmabar^\nu \sigma^\rho \, \xibar + i \,  \partial_\nu (\xi \partial_\mu \pi) \, \sigma^\mu  \sigmabar^\nu \sigma^0 \, \xibar\Big) + {\rm h.c.}
\ee
Using $\vevphi = H \Mpl \sqrt{2\varepsilon}$ and $\Lambda = \sqrt{\Mpl H}$, we can express the coefficient of this operator in terms of inflationary parameters,
\be
\label{eq:leading3pt}
\mathcal{L}_{\text{3-pt}} = \frac{1}{2\sqrt{2} \Lambda^2} \frac{\sqrt{\varepsilon}}{3 - \varepsilon} \Big(i \, \partial_\rho \pi \, \partial_\nu \xi \, \sigma^0  \sigmabar^\nu \sigma^\rho \, \xibar + i \,  \partial_\nu (\xi \partial_\mu \pi) \, \sigma^\mu  \sigmabar^\nu \sigma^0 \, \xibar\Big) + {\rm h.c.}
\ee
As expected from \Eq{eq:naturalcutoff}, this dimension-6 operator is naturally suppressed by the cutoff scale $\Lambda$ instead of $\Mpl$, and it includes the expected slow-roll suppression factor $\sqrt{\varepsilon}$. Similarly, the term where both factors of $\partial \varphi$ are replaced by $\partial \pi$ leads to 4-point interactions,
\be
\label{eq:leading4pt}
\mathcal{L}_{\text{4-pt}} = \frac{i}{12 \Lambda^4 (3-\varepsilon)} \partial_\rho \pi (\partial_\mu \pi \, \partial_\nu \xi + \partial_\mu \partial_\nu \pi \, \xi) \sigma^\mu \sigmabar^\nu \sigma^\rho \xibar + {\rm h.c.}
\ee
This dimension-8 operator is suppressed by $1/\Lambda^4$ as expected, but is not slow-roll suppressed to leading order. Both of these vertices were shown schematically in \Fig{fig:3and4pt}.

\section{Supergravity EFT of inflation}
\label{sec:SUGRA}
Up to this point, we have worked out the leading-order effects of our theory in the rigid SUSY limit ($\Mpl\to\infty$). This let us focus on the goldstone and goldstino fluctuations without worrying about the complications inherited from the inflating FRW background. In this limit, we were still able to see the reduced speed of sound of the goldstino and derive the leading goldstino-goldstone interactions of the EFT.

In this section, we work out the minimal SUGRA EFT of inflation by coupling our nonlinear chiral multiplets $\ANL$ and $\XNL$ to SUGRA. The goldstino and gravitino become fluctuations on an FRW background with metric
\be
ds^2 = -dt^2 + a(t)^2 d\vec{x}^2, \label{FRW-flat-slicing}
\ee
where $a(t) = e^{Ht}$ is the scale factor.  As in the EFT of inflation, the relationship between the FRW parameters $H$ and $\dot{H}$ and the components of the inflaton stress-energy tensor is set dynamically by the Friedmann equations. In addition to having a dynamical metric, SUGRA introduces the spin-$3/2$ gravitino field as the superpartner to the graviton. Just as the metric eats the inflaton fluctuation in unitary gauge, the gravitino eats the goldstino via the super-Higgs mechanism, leading to a spin-$1/2$ longitudinal mode. We will find that the mode equation for the spin-$1/2$ mode will reduce to precisely that of the goldstino computed in the rigid limit, giving a speed of sound $c_s = 1 - 2\varepsilon/3$.

Additionally, we allow a negative cosmological constant $\Lambda_0 = -3\Mpl^2 /\ell_{\rm AdS}^2$. This term gives the gravitino a SUSY-preserving mass in pure SUGRA, with $|m_{3/2}| = 1/\ell_{\rm AdS}$. As discussed in \Sec{Sec:scales}, goldstino equivalence no longer holds for large $|m_{3/2}|$, and we cannot ignore the couplings to the transverse polarizations of the gravitino in the EFT. In this section, we depart from the goldstino equivalence limit and allow the case of $|m_{3/2}| \gtrsim H$. This hierarchy of scales is interesting because in the case where $|m_{3/2}|$ is the same during and after inflation, it corresponds to extremely high-scale SUSY breaking, where masses of superpartners today are well above the Hubble scale during inflation. In this case, superpartners would almost certainly be unobservable at terrestrial colliders, but could still give observable signatures during inflation \cite{Craig:2014rta}.

We work in the chiral superspace formalism, where the minimal SUGRA action is  
\be
S = \Mpl^2 \int d^4x\ d^2\Theta\ 2 \mb{\mathcal{E}}\ \Big\{ \frac{3}{8} (\overline{\mathcal{D}}^2 - 8\mb{\mathcal{R}})e^{-\mb{K}[\XNL, \XNL^\dagger, \BNL]/3\Mpl^2}   + \frac{1}{\Mpl^2} \mb{W}[\XNL]\Big\} + {\rm h.c.}, \label{S-SUGRA-general}
\ee
where $\mb{K}[\XNL, \XNL^\dagger, \BNL]$ is the \kahler potential, $\mb{W}[\XNL]$ is the superpotential, and $\mb{\mathcal{E}}$, $\mb{\mathcal{R}}$, and $\mathcal{D}$ are the chiral density, chiral superspace curvature, and superspace covariant derivative, respectively.\footnote{The expressions for $\mb{\mathcal{E}}$ and $\mb{\mathcal{R}}$ will not be needed here, but are written out in Eq.~(21.8) of \Ref{wess1992supersymmetry}.}  Additionally, $\Theta$ is the chiral superspace variable, which carries local Lorentz indices. 

\subsection{Linear \kahler frame}
\label{sec:linearframe}

The most general \kahler potential and superpotential without derivatives were found previously in \Eq{eq:generalLag}, which we repeat here for convenience:
\begin{align}
\mb{K}[\XNL, \XNL^\dagger, \BNL] & = \XNL^\dagger \XNL + \BNL^2 + \alpha \BNL + \beta \XNL + \beta^* \XNL^\dagger,\nonumber \\
\mb{W}[\XNL] & = \Mpl^2 |m_{3/2}| + f\XNL, \label{super-potential}
\end{align}
where we have added a constant term to the superpotential to give an additional cosmological constant $\Lambda_0$.  Higher-order polynomial terms in $\XNL$ and $\BNL$ in the \kahler potential vanish by the constraints $\XNL^2 = \BNL^3 = \XNL \BNL = 0$. Because we are working in the chiral superspace formalism, \Eq{super-potential} should be understood as depending on the chiral superfield $\ANL$, which can be found by substituting \Eq{eq:BNL-def}.

In the superpotential, $|m_{3/2}| =  |\langle \mb{W} \rangle|/\Mpl^2$ is the gravitino mass in pure SUGRA. We recall the standard result from SUGRA that the contribution to the vacuum energy from the superpotential is nonpositive, so the positive vacuum energy during inflation must come from the \kahler potential. Note that for nonzero $m_{3/2}$, it is no longer true that $f =  \Mpl H\sqrt{3 - \varepsilon}$ as there is an additional contribution to $H$ in the Friedmann equations from the cosmological constant $\Lambda_0$.

The term proportional $\beta$ in \Eq{super-potential} can be removed by a \kahler transformation, which reshuffles terms between the \kahler potential and the superpotential as
\be
\mb{K} \to \mb{K} + {\bf \Omega} + {\bf \Omega}^\dagger, \qquad \mb{W} \to e^{-{\bf \Omega}/\Mpl^2}\mb{W},
\label{eq:KahlerTransform}
\ee
where the \kahler transformation parameter $\bf \Omega$ is a (dimension-2) chiral superfield.\footnote{Since our EFT has no gauge multiplets, there is no \kahler anomaly to worry about.} Taking ${\bf \Omega} = -\beta \XNL$ removes the terms proportional to $\beta$ from \Eq{super-potential}, but it adds no new terms to the superpotential thanks to the constraint $\XNL^2 = 0$ and it instead just shifts $f \to f + \beta |m_{3/2}|$. Thus we can work in the K\"{a}hler-transformed frame and absorb the effects of $\beta$ into a redefinition of $f$. 

After removing $\beta$, the \kahler potential and superpotential are
\begin{align}
\mb{K}[\XNL, \XNL^\dagger, \BNL]  & = \XNL^\dagger \XNL + \BNL^2 + \alpha \BNL, \nonumber \\
 \mb{W}[\XNL] &  = \Mpl^2 |m_{3/2}| + f\XNL.
 \label{eq:KLinear}
\end{align}
We refer to this as the ``linear'' \kahler frame due to the presence of the linear term $\alpha \BNL$. From \Eq{eq:KLinear}, we can see that $\alpha$ is technically natural in the sense of 't Hooft, since when $\alpha = 0$ the \kahler potential has the enhanced $\mathbb{Z}_2$ symmetry $\BNL \to - \BNL$, or $\partial_\mu \pi \to -\partial_\mu \pi$ at the Lagrangian level. Thus $\alpha$ may be expected to be small, but as a new parameter it is still worthwhile to investigate its effects.

In this linear \kahler frame, the vector auxiliary field $b_\mu$ obtains a vev in the time component $b_0$.  As shown in \App{app:Linear}, the value of the vev is
\be
\label{eq:b0advert}
\langle b_0 \rangle = \frac{i\alpha}{2\Mpl^2} \vevphi \qquad \Big(\langle \dot{\mb{W}} \rangle = 0\Big).
\ee
The Lorentz-violating $b_0$ vev gives rises to a Lorentz-violating gravitino bilinear,
\be
\mathcal{L}_{3/2} \supset \frac{i \alpha}{4\Mpl^2} \vevphi \epsilon^{0 \mu \nu \kappa}\psibar_\mu \sigma_\nu  \psi_\kappa.
\label{eq:LVmass}
\ee
This establishes the connection between $\langle b_\mu \rangle$ and Lorentz-breaking of SUSY as suggested in \Ref{Baumann:2011nk}.\footnote{Note that $\langle M \rangle$, the vev of the SUGRA scalar auxiliary field, is nonvanishing.  Our formalism differs from that of \Ref{Baumann:2011nk} by the addition of the $R$-symmetry breaking parameter $|m_{3/2}|$ (see \App{app:Linear}).}  We discuss the role of this $\alpha$ term in \Sec{sec:alpharole}.

\subsection{Canonical \kahler frame}
\label{sec:kahler}
 
Using another \kahler transformation, we can also remove the linear term $\alpha \BNL$ from the \kahler potential, at the expense of making the superpotential non-polynomial. Taking ${\bf \Omega} = \frac{i\alpha}{2}\ANL$, the \kahler transformation of \Eq{eq:KahlerTransform} gives
\begin{align}
\mb{K} & = \XNL^\dagger \XNL + \BNL^2, \nonumber \\
\mb{W} & = e^{-i\alpha \ANL/2\Mpl^2 }\Mpl^2 |m_{3/2}| + f\XNL.
\label{eq:Kcanonical}
\end{align}
Here, we have used the result from \Eq{eq:Xredef} that allows us to absorb any $\ANL$-dependent phase of $f$ into a redefinition of $\XNL$.  We refer to this as the ``canonical'' \kahler frame.\footnote{In the minimal SUGRA formalism, there is no kinetic mixing between the gravitino and the goldstino even for nonzero background values of the auxiliary fields, so the choice of canonical frame or linear frame is simply one of convenience.  See \Ref{Cheung:2011jp}, however, for a discussion of frame-dependent subtleties which arise in the conformal compensator formalism.}

At first glance, the phase in $\mb{W}$ seems to break the manifest shift symmetry we had previously imposed by writing the \kahler potential as a function of $\BNL$ only.\footnote{Note that $\ANL \XNL \neq 0$ and $\ANL$ is not nilpotent in contrast to $\XNL$, so the exponential prefactor is non-trivial.}  However, the lowest component of $\ANL$ does not obtain a vev as long as we assume there is no Lorentz-breaking goldstino condensate (see \Eq{eq:Blowest}), so the vev of the superpotential still respects the shift symmetry, $\langle \mb{W} \rangle = \Mpl^2 |m_{3/2}|$. Instead, the effect of $\alpha$ is to give $\dot{\mb{W}}$ a nonzero imaginary vev,
\be
\langle \dot{\mb{W}} \rangle = -\frac{i\alpha}{2\Mpl^2}\vevphi |m_{3/2}| \qquad \Big(\langle b_\mu \rangle = 0\Big).
\ee
We emphasize that in the canonical \kahler frame, the vector auxiliary field of SUGRA $b_\mu$ does not obtain a Lorentz-violating vev. $\langle \dot{\mb{W}} \rangle$ can be interpreted as time-dependent phase of the gravitino mass,
\be
\label{eq:m32phase}
m_{3/2} = e^{-i \alpha \vevphi t/2\Mpl^2}\,|m_{3/2}|.
\ee 
This can be understood as a local (in time) redefinition of the left- and right-handed Weyl spinors which are coupled through the mass term, a point of view we will return to in \Sec{sec:alpharole}.

Alternatively, one can remove the time-dependent mass phase with a field redefinition of the gravitino. Consider the unitary gauge Rarita-Schwinger Lagrangian,
\be
\mathcal{L}_{3/2} = \epsilon^{\mu \nu \rho \kappa}\psibar_\mu \sigma_\nu D_\rho \psi_\kappa - m_{3/2}\psi_\mu \sigma^{\mu \nu} \psi_\nu - m_{3/2}^* \psibar_\mu \sigmabar^{\mu \nu} \psibar_\nu,
\ee
where $D_\rho$ is the covariant derivative for a gravitino.  The field redefinition 
\be
\psi_\mu \to e^{i \alpha \varphi/4\Mpl^2}\,\psi_\mu, \qquad \psibar_\mu \to e^{-i \alpha \varphi/4\Mpl^2}\,\psibar_\mu
\label{eq:psiredef}
\ee
restores a real and time-independent $m_{3/2} = |m_{3/2}|$. However, the Rarita-Schwinger kinetic term now generates an extra term proportional to $\epsilon^{0 \mu \nu \kappa}\psibar_\mu \sigma_\nu  \psi_\kappa$, which is (not surprisingly) the same Lorentz-violating term from \Eq{eq:LVmass} we found in the linear \kahler frame.

\subsection{The role of $\alpha$}
\label{sec:alpharole}

The presence of the parameter $\alpha$ can potentially lead to interesting phenomenological consequences.  From \Eq{eq:m32phase}, the gravitino obtains a Lorentz-violating, time-dependent phase proportional to $\alpha$ (or equivalently, the bilinear term in \Eq{eq:LVmass}). For a fermion with a Dirac mass, a $\vevphi$-dependent mass phase would source a chemical potential for the fermion in the inflationary background, similar to \Refs{Cohen:1987vi,Cohen:1988kt,Hook:2014mla,DAgnolo:2015pha}.  For a fermion with a Majorana mass, the effect of this phase is somewhat less intuitive, so in \App{app:alphadispersion} we work out a toy example with a Majorana spin-$1/2$ fermion before deriving the phase-modified dispersion relations for a Rarita-Schwinger field.  In addition, there are 3- and 4-point vertices arising from \Eq{eq:LVmass} and related terms, so scattering amplitudes and loop diagrams will contain contributions proportional to $\alpha$.

That said, effects proportional to $\alpha$ are expected to be extremely small.  Parametrically, the coefficient in \Eq{eq:LVmass} is
\be
\frac{\alpha\vevphi}{4\Mpl^2}  = \frac{\alpha}{2\sqrt{2} \Mpl}\sqrt{\varepsilon} H.
\label{eq:alphasmall}
\ee
Since $\alpha < \Lambda$ for the validity of the EFT and $\varepsilon \ll 1$ during slow-roll inflation, the effects of $\alpha$ are doubly suppressed compared to $H$. In the case where $|m_{3/2}|$ is of order $H$, the effects of the expanding universe (\ie\ Hubble friction) will dominate the effects of $\alpha$.  In the opposite limit where $|m_{3/2}|$ can be neglected, $\alpha$ is unphysical.  This fact is clear in the canonical \kahler frame (without redefining the gravitino), since $\alpha$ only appears as a phase in the gravitino mass, which is therefore irrelevant when $|m_{3/2}| \to 0$.  Thus, in the EFT of inflation, there is no parametric regime where $\alpha$ gives the leading corrections to the rigid flat space mode equations.  Apart from an interesting calculation in \App{app:alphadispersion-RS}, we will set $\alpha=0$ from now on, though we still leave $|m_{3/2}|$ arbitrary.

\subsection{Revisiting the slow gravitino}
\label{sec:gravdispersion}

Having dispensed with $\alpha$, the leading-order SUGRA effects arise from the gravitino equations of motion.  To derive the mode equations, it is simplest to perform this analysis in unitary gauge, where the goldstino is eaten by the gravitino.  With $\alpha = 0$, the unitary-gauge gravitino Lagrangian is
\be
\mathcal{L}_{3/2} = \epsilon^{\mu \nu \rho \kappa}\psibar_\mu \sigma_\nu D_\rho \psi_\kappa  - |m_{3/2}|\left (\psi_\mu \sigma^{\mu \nu} \psi_\nu + \psibar_\mu \sigmabar^{\mu \nu} \psibar_\nu \right).
\label{eq:RSalpha}
\ee
With a real mass, we can combine the two Weyl spinors $\psi_\mu$ and $\psibar_\mu$ into a 4-component Majorana spinor $\Psi_\mu$, decoupling the equations of motion for $\Psi_\mu$ and $\overline{\Psi}_\mu \equiv \Psi_\mu^\dagger \gamma^0$.  Furthermore, since the Lagrangian coefficients are manifestly time-independent, the identification of the propagating polarization spinors is straightforward.

In fact, \Ref{Giudice:1999yt} already worked out the gravitino equations of motion in an FRW background in four-component fermion notation, so we can simply recall the results here.\footnote{For similar work along these lines, see \Refs{Maroto:1999ch,Giudice:1999am,Kallosh:1999jj,Kallosh:2000ve}.}   After considering general $|m_{3/2}|$, we will examine the $|m_{3/2}| \to 0$ limit to see how the gravitino equations of motion map onto the goldstino modified speed of sound result from \Sec{sec:kinetic}.

First, it is convenient to write the FRW metric in conformal time,
\be
ds^2 = a^2(\eta)(-d\eta^2 + d\vec{x}^2),
\ee
with $\eta$ defined by the relation $dt = a(t(\eta)) d \eta$. In terms of conformal time, the FRW parameters are given by
\be
H = \frac{a'}{a^2}, \qquad
\dot{H} + 2H^2 = \frac{a''}{a^3}, \label{eq:conformal-FRW}
\ee
where primes denote derivatives with respect to conformal time, $a' = da/d\eta$.  We also make use of the vierbein
\be
e^\mu_a = a^{-1} \delta^\mu_a
\ee
and curved-space objects
\be
\hat{\gamma}^\mu \equiv e_a^\mu \gamma^a, \qquad \epsilon^{\mu \nu \rho \sigma} \equiv e e_a^\mu e_b^\nu e_c^\rho e_d^\sigma \epsilon^{abcd}.
\ee
Here, Latin indices denote tangent-space indices and Greek indices are space-time indices.

Next, we rewrite the equation of motion for $\Psi_\sigma$ as the condition that a 4-vector of spinors $R^\mu$ vanish identically
\be
R^\mu \equiv \epsilon^{\mu \nu \rho \sigma} \gamma_5 \hat{\gamma}_\mu \hat{D}_\rho \Psi_\sigma  = 0,
\label{eq:RRS}
\ee
where the covariant derivative
\be
\hat{D}_\mu = \partial_\mu - \frac{a'}{4a^2}\left( \hat{\gamma}_\mu\gamma^0 - \gamma^0\hat{\gamma}_\mu\right) -  \frac{i}{2} |m_{3/2}| \hat{\gamma}_\mu
\ee
is defined to include the mass term. We can use the spatial isometries of the FRW metric to Fourier transform the spatial dependence of the gravitino field,
\be
\Psi^\mu(\eta, \vec{x}) \sim \int d^3 k \, e^{i\vec{k}\cdot\vec{x}}\Psi^\mu_{\vec{k}}(\eta).
\ee
The Rarita-Schwinger equation (\ref{eq:RRS}) implies two algebraic constraints on $\Psi^\mu_{\vec{k}}$. The equation for $R^0$ contains no time derivatives thanks to the antisymmetry of the Levi-Civita tensor, and so gives an algebraic constraint relating the components $\Psi^i_{\vec{k}}$, $i = 1, 2, 3$. 
The second constraint comes from operating on \Eq{eq:RRS} with $\hat{D}_\mu$. \Ref{Giudice:1999yt} showed that one can solve for $\Psi^0$ as
\be
\Psi^0_{\vec{k}} = c \sum_{i=1}^3 \hat{\gamma}_i \Psi^i_{\vec{k}},
\label{eq:PsiCond}
\ee
with $c$ a matrix in spinor space which reduces to $c = \gamma^0$ in flat space, recovering the familiar Rarita-Schwinger constraint $\gamma \cdot \Psi = 0$.

Using these constraints, one can work out the mode equations for each of the polarizations \cite{Giudice:1999yt}.   The mode equations for the spin-1/2 component $\Psi_{1/2, \vec{k}}$ and the spin-3/2 component $\Psi_{3/2, \vec{k}}$ (which are linear combinations of the two spinors remaining after solving the two constraints) are
\begin{align}
& \left [i \gamma^0 \partial_0 + i \frac{5a'}{2a}\gamma^0- |m_{3/2}|a + (A + iB \gamma^0) \vec{k}\cdot\vec{\gamma}\right]\Psi_{1/2, \vec{k}} = 0, \label{gravitino-mode-spin-1/2} \\
& \left [i \gamma^0 \partial_0 + i \frac{5a'}{2a}\gamma^0- |m_{3/2}|a + \vec{k}\cdot\vec{\gamma}\right]\Psi_{3/2, \vec{k}}= 0, \label{gravitino-mode-spin-3/2} 
\end{align}
with
\begin{align}
A & = \frac{1}{3\left(\frac{(a')^2}{a^4} + |m_{3/2}|^2\right)^2}\left[ 2 \frac{a''}{a^3}\left(|m_{3/2}|^2 - \frac{(a')^2}{a^4}\right) + \frac{(a')^4}{a^8}
 -4 |m_{3/2}|^2 \frac{(a')^2}{a^4} + 3|m_{3/2}|^4 \right], \\
B & = \frac{2|m_{3/2}|}{3\left(\frac{(a')^2}{a^4} + |m_{3/2}|^2\right)^2} \left [ 2 \frac{a''a'}{a^5} - \frac{(a')^3}{a^6} + 3|m_{3/2}|^2\frac{a'}{a^2} \right].
\end{align}
We see that $\Psi_{3/2}$ has a canonical kinetic term, but the spatial part of the kinetic term for $\Psi_{1/2}$ is modified by the coefficients $A$ and $B$.  The term proportional to $a'/a$ is the Hubble friction term, which affects all four polarizations equally.

The above results are valid for any value of $|m_{3/2}|$ in any FRW background.  This means that we can perform a third (and final) cross check of the modified speed of sound result of \Sec{sec:kinetic} by isolating the longitudinal spin-$1/2$ polarization in the $|m_{3/2}| \to 0$ limit.  Looking at modes $\omega > H$ to compare to flat space, we find that $\Psi_{1/2, \vec{k}}$ has a dispersion relation $\omega/k = |A|$, with
\be
A = \frac{1}{3} - \frac{2a'' a}{3(a')^2}.
\ee
Plugging in using \Eq{eq:conformal-FRW} and the definition of the slow roll parameter, we find
\be
c_s = |A| = 1 - \frac{2}{3}\varepsilon = \frac{1 - \kappa}{1+\kappa}, \label{gravitino-small-cs}
\ee
exactly reproducing our earlier results obtained from the goldstino equivalence regime.

It is intriguing that the nontrivial speed of sound is here a consequence of the FRW background, whereas in \Sec{sec:kinetic} it was due to the nontrivial vev of the inflaton energy-momentum tensor, evaluated on a flat background. This is a direct consequence of SUSY and the structure of SUGRA. In the rigid flat limit, $\langle T_{\mu \nu} \rangle$ cannot back-react on the geometry, but due to the Ward-Takahashi identity for the supercurrent, it is $\langle T_{\mu \nu} \rangle$ which determines the goldstino dispersion relation. For finite $\Mpl$, $\langle T_{\mu \nu} \rangle$ sources the curvature $R$ of the background, and the gravity multiplet containing both $R$ and $\psi_\mu$ communicates the nontrivial goldstino dispersion to the longitudinal component of the gravitino through the super-Higgs mechanism. Finally, note that in flat space with unbroken SUSY ($m_{3/2} = 0$ and $a = 1$), $A  = B = 0$.  In this limit, the spin-1/2 mode does not propagate, and we recover the two physical polarizations appropriate for unbroken SUSY.

\subsection{Loop corrections to inflationary observables} 

It would be rather surprising if there were large observable consequences of the goldstino of SUSY inflation. Cosmologically, fermions are only observable through loop effects. Loop corrections in inflation were explored in \Ref{Weinberg:2005vy} for reasons of considering (even potentially unobservable) theoretical consequences of the theory. This was followed up in \Refs{Senatore:2009cf,Senatore:2012nq,Pimentel:2012tw}, in which the authors investigated an unphysical logarithmic running claimed in \Ref{Weinberg:2005vy}, and also recently revisited in \Refs{Senatore:2012ya,Assassi:2012et,Green:2013rd,Arkani-Hamed:2015bza}.

The main cosmological observable is the dimensionless curvature mode $\zeta$, which in the EFT of inflation is proportional to $\pi$ \cite{Cheung:2007st},
\be
\zeta = -\frac{H}{\vevphi}\pi.
\ee
The simplest observable is the two-point function, the form of which is fixed by conformal symmetry. In momentum space, this is given by
\be
\langle \zeta \zeta \rangle \sim \frac{1}{k^{3-n_s}}, \label{eq:two-point-function}
\ee
where $n_s=-4\varepsilon$ is the spectral tilt.\footnote{In this section, as in the rest of the paper, we have set the second slow roll parameter $\eta$ to zero.} In the de Sitter limit ($\varepsilon=0$), conformal invariance is exact and \Eq{eq:two-point-function} is the two-point function for a three-dimensional conformal operator dual to a massless scalar in de Sitter space. In the slow roll limit ($\varepsilon>0$), $n_s$ characterizes the deviation from exact scale invariance.

\begin{figure}[t]
\begin{center}
\includegraphics[width=0.9\columnwidth]{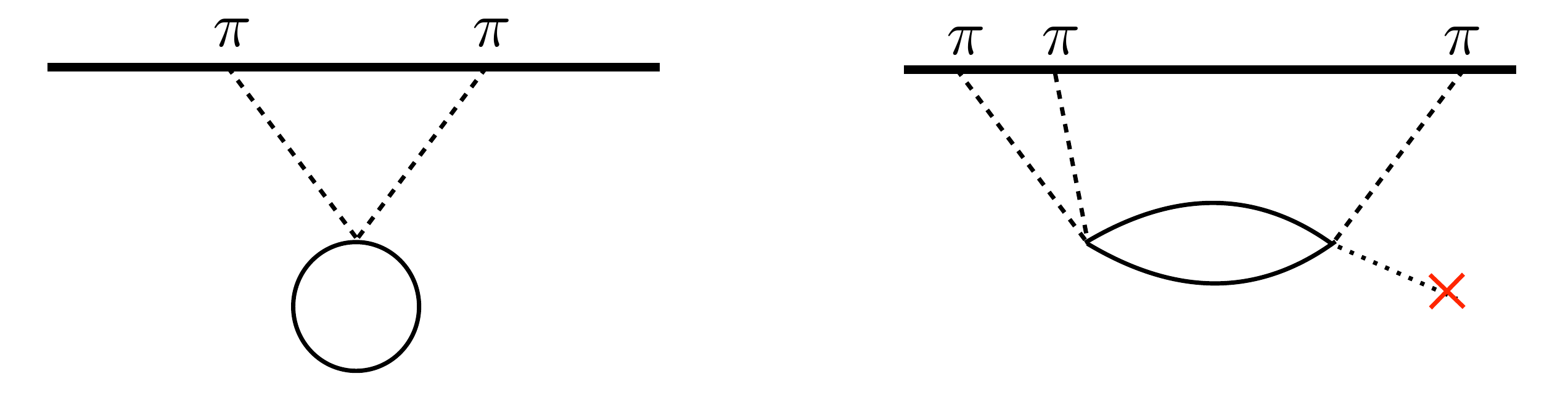}
\caption{Leading diagrams with loops of goldstinos. {\bf Left:} Contribution to the goldstone two-point function. {\bf Right:} Contribution to the goldstone three-point function.}
\label{fig:2-3pt}
\end{center}
\end{figure}

In our EFT, there will be corrections to the two point function from goldstino loops shown in the left side of \Fig{fig:2-3pt}.  In general, such ``self-energy'' type corrections compute the anomalous dimensions of the three-dimensional conformal operator dual to the the inflaton fluctuation.  The leading corrections from goldstino loops will be proportional to the 4-point $\pi \pi \xi \xi$ vertex, which is given by \Eq{eq:leading4pt} and scales like $1/\Lambda^4$. By dimensional analysis (\ie\ using factors of $H$ from the derivatives in the interaction vertices to make up the dimensions in the numerator), the size of loop corrections is expected to be \cite{Senatore:2009cf,Senatore:2012nq,Pimentel:2012tw,Senatore:2012ya}
\be
\langle \zeta \zeta \rangle \sim \frac{1}{k^{3-n_s}}\bigg(1+ \#\frac{H^4}{\Lambda^4} \log \frac{k}{a \mu} \bigg) ,
\ee
where we have thrown away an expected UV-divergence \cite{Anninos:2014lwa}, and the exact relation will be scheme-dependent. Here $\mu$ is the renormalization scale, and the scale factor $a$ is implicitly determined by $k/a = H$, where all quantities are evaluated at horizon crossing.\footnote{We thank Tarek Anous for a discussion of this point and the anonymous referee for further clarifications.} However, the form of the answer is constrained due to conformal symmetry and so we expect this to be expressed as \cite{Marolf:2010zp,Krotov:2010ma}
\be
\langle \zeta \zeta \rangle \sim \frac{1}{k^{3-n_s+2\gamma}}, \label{eq:two-point-function-shift}
\ee
where $\gamma \sim H^4 / \Lambda^4  = H^2 /\Mpl^2$ is the shift due to the goldstino loop. This highlights a bug (or feature) of the two-point function: the symmetry is so constraining that we cannot distinguish between the effects of the gravitino and effects due to inflaton self-interactions or gravitational effects (for example, graviton loops). The form of the two-point function is restricted to be a power law decay in $k$ and can have no other features. 

A recent program in inflation attempts to exploit the high energy scale $H$ set by inflation and use cosmological observations to provide evidence for new particles.  This was first done in the context of indirect effects on the goldstone $\pi$ \cite{Chen:2009zp}, and in many other scenarios by \Refs{Senatore:2010wk,Baumann:2011nm,Assassi:2013gxa,2012JCAP...08..033S,2012JCAP...08..019N,Craig:2014rta,Arkani-Hamed:2015bza,Kehagias:2015jha}. Such observables require going beyond the two-point function. In de Sitter space, the three-point function at late times is also highly constrained by conformal symmetry, and the first nontrivial correlation function is the four-point function. In inflation, the three-point function can be thought of as a de Sitter four-point function with one of the legs set to the inflationary background.
  
The leading goldstino contribution to the inflaton three-point function is shown in the right side of \Fig{fig:2-3pt}.   As emphasized in \Ref{Arkani-Hamed:2015bza}, only new particles can lead to non-analyticities in the three-point function. These effects are necessarily nonperturbative in the $H/\Mpl$ power counting, so cannot be detected in the usual power-expanded EFT of inflation.\footnote{We thank Nima Arkani-Hamed for a discussion of these points.}  In our SUSY EFT, however we have two low energy modes, the goldstone and the gravitino.  Therefore, even though goldstone self-interactions are expanded in powers of $H/\Mpl$ in the EFT, we still expect to see resonances of the gravitino in the three-point function of the goldstone.  

In general, loop contributions to the goldstone three-point function will depend on properties of the particle in the loop, such as its mass and spin. \Ref{Arkani-Hamed:2015bza} identifies two different observable signatures of a new particle. For heavy particles $m \gtrsim H$, the correlation function exhibits an oscillatory interference effect.  This signature would be present for gravitino loops if $m_{3/2}$ is comparable to $H$.  For light particles, the correlation function exhibits a power-law decay with the exponent sensitive to the mass of the particle.  This is the case for $m_{3/2} \ll H$, where we expect inflationary observables to depend primarily on the goldstino (equivalently, longitudinal gravitino).  Optimistically, there might even be observable consequences from the different dispersions of the two gravitino modes (\Eqs{gravitino-mode-spin-1/2}{gravitino-mode-spin-3/2}).\footnote{Of course, there will also be contributions from goldstone loops, but one may hope that these could be disentangled from the goldstino/gravitino loops because of the different spins of these particles.}  We leave a detailed calculation of these effects to future work.

\section{Conclusion}
\label{sec:conclusion}

In this paper, we have established that the minimal low energy degrees of freedom in a consistent SUSY model of inflation are a real scalar goldstone $\pi$ and a Weyl fermion goldstino $\xi$. These correspond to the two symmetries spontaneously broken by the inflationary background: time translation invariance and SUSY. Realistic models may of course have other states between the IR cutoff $H$ and the UV cutoff $\Lambda = \sqrt{\Mpl H}$, such as the extra scalar emphasized in \Ref{Baumann:2011nk}, but $\pi$ and $\xi$ are the irreducible degrees of freedom present in every SUSY inflationary scenario. 

Our theory can be regarded as the minimal SUSY extension of the EFT of inflation of \Ref{Cheung:2007st}, in the context of slow-roll inflation.  By using nonlinear multiplets, we can express $\pi$  and $\xi$ compactly within two constrained superfields, $\XNL$ and $\BNL$. We have also shown how to relate our theory to the SUSY EFT studied in \Ref{Baumann:2011nk}, which includes the scalar partner of the inflaton and an extra fermion in the inflaton multiplet. We recovered the nonlinear constraints on $\BNL$ by pushing up the mass of the extra states above the EFT scale and integrating them out.

This theory lets us compute the model-independent irreducible SUSY contribution to inflationary observables in the slow-roll approximation.  Of course, this formalism predicts that there will probably be no such observations in the near future, since the loop-induced effects of the goldstino are quite small. We expect the one-loop contribution of the goldstino to the scalar two-point function to contribute parametrically as $H^2/\Mpl^2$ compared to the tree level contribution, which is extremely small unless $H$ is very large.  Similarly, the contribution of goldstino loops to the scalar three-point function will carry signatures of the mass and spin of the goldstino, but these will be slow-roll, power-law, and possibly even exponentially suppressed. Nevertheless, it could be interesting to actually compute the goldstino one-loop contribution by continuing a perturbative calculation in EAdS space to de Sitter space as in \Refs{Maldacena:2002vr,Ghosh:2014kba,Kundu:2014gxa, Anninos:2014lwa}.  Also, it is worth noting that goldstino dynamics might be relevant for (p)reheating (see e.g.~\cite{Nilles:2001ry,Giudice:1999yt,Giudice:1999am,Kallosh:1999jj,Kallosh:2000ve}), though that era is strictly speaking outside of the range of validity of the present EFT.  

Finally, we speculate about the effects of extremely high-scale SUSY breaking. If $|m_{3/2}|$ does not change during or after inflation, the gravitino mass during inflation is related to the scale of SUSY-breaking today. After coupling our EFT of inflation to SUGRA, we have identified an additional free parameter $\alpha$, parameterizing a time-dependent phase for the gravitino mass $m_{3/2}$. If $|m_{3/2}| > H$, there are additional interaction vertices proportional to the new parameter $\alpha$. The effects of $\alpha$ are parametrically both Planck- and slow-roll suppressed, but nevertheless calculable with our EFT. Since this is also the region of parameter space where the scale of SUSY-breaking is so high as to never be directly observable at terrestrial colliders, one can hope that the universe can be used as a ``cosmological collider'' \cite{Cheung:2007st,Senatore:2010wk,Chen:2009zp,Baumann:2011nm,Assassi:2013gxa,2012JCAP...08..033S,2012JCAP...08..019N,Craig:2014rta,Arkani-Hamed:2015bza,Kehagias:2015jha} to probe the effects of SUSY in the early universe.  We leave an investigation of these potentially interesting effects to future work.

\acknowledgments{We thank Allan Adams, Tarek Anous, Nima Arkani-Hamed, Cliff Cheung, Liam Fitzpatrick, Dan Freedman, Daniel Green, Anupam Mazumdar, Leonardo Senatore, Julian Sonner, Lorenzo Sorbo, Douglas Stanford, and Zachary Thomas for helpful conversations.  This work is supported by the U.S. Department of Energy (DOE) under cooperative research agreement Contract Number DE-SC00012567.  Y.K.\ is also supported by an NSF Graduate Fellowship. D.R.\ is also supported by the Fannie and John Hertz Foundation.  J.T.\ is also supported by the DOE Early Career research program DE-SC0006389 and by a Sloan Research Fellowship from the Alfred P. Sloan Foundation.}

\appendix
\section{Components of $\BNL$}
\label{app:BNL}
The ``axion'' superfield $\ANL$ of \Ref{Komargodski:2009rz} is a chiral superfield whose lowest component is
\begin{align}
\ANL |  \equiv a & =  \varphi + \frac{i}{2|F|^2}\xi \sigma^\mu \xibar \partial_\mu \varphi + \frac{1}{8|F|^4} \left (\xi^2 \partial_\nu \xibar \sigmabar^\mu \sigma^\nu \xibar  - {\rm c.c.} \right)\partial_\mu \varphi \\ \nonumber
& -\frac{i}{32|F|^6}\xi^2 \xibar^2 \partial_\mu \xibar (\sigmabar^\rho \sigma^\mu \sigmabar^\nu + \sigmabar^\mu \sigma^\nu \sigmabar^\rho)\partial_\nu \xi \partial_\rho \varphi.
\end{align}
In particular, ${\rm Re}( \ANL|)= \varphi$. The full superfield $\ANL$ has components
\be
\ANL = a + \frac{i\sqrt{2}}{\overline{F}} \theta \sigma^\mu \xibar \partial_\mu a + \frac{1}{\overline{F}^2}\theta^2 \left (-\partial_\nu \xibar \sigmabar^\mu \sigma^\nu \xibar \partial_\mu a + \frac{1}{2}\xibar^2 \partial^2 a \right),
\label{eq:Afull}
\ee
where all fields are understood to be functions of $y^\mu = x^\mu + i\theta \sigma^\mu \thetabar$.\footnote{In SUGRA, $\partial_\mu \xi$ should be understood to mean $\nabla_\mu \xi$.}

We define the real superfield $\BNL = \frac{1}{2i}(\ANL - \ANL^\dagger)$ as in \Eq{eq:BNL-def}, whose components can then be obtained by Taylor-expanding \Eq{eq:Afull} in $y$. We will be primarily interested in the components of $\BNL$ at most quadratic in $\xi$ and $\xibar$:
\begin{align}
\label{eq:Blowest}
\BNL| & = \frac{1}{2|F|^2}\xi \sigma^\mu \xibar \partial_\mu \varphi, \\
\label{eq:Btheta}
\BNL|_\theta & = \frac{1}{F\sqrt{2}}  \sigma^\mu \xibar \partial_\mu \varphi  , \\
\label{eq:Bthetasq}
\BNL|_{\theta^2} & = \frac{i}{2\overline{F}^2}\left (\partial_\nu \xibar \sigmabar^\mu \sigma^\nu \xibar \partial_\mu \varphi - \frac{1}{2}\xibar^2 \partial^2 \varphi \right), \\
\BNL|_{\theta \sigma^\mu \thetabar} & = \partial_\mu \varphi, \\
\label{eq:Bthetasqthetabar}
\BNL|_{\theta^2 \thetabar} & = \frac{i}{2F\sqrt{2}} \sigmabar^\nu \sigma^\mu \left (\partial_\nu \xibar \partial_\mu \varphi + \xibar \partial_\mu \partial_\nu \varphi \right), \\
\BNL|_{\theta^2 \thetabar^2} & = \frac{1}{4|F|^2} \square \left (\xi \sigma^\mu \xibar \partial_\mu \varphi \right),
\end{align}
where the $\thetabar$, $\thetabar^2$, and $\thetabar^2 \theta$ components are obtained by complex-conjugating \Eqss{eq:Btheta}{eq:Bthetasq}{eq:Bthetasqthetabar} respectively.

\section{Lorentz breaking and auxiliary fields}
\label{app:Linear}
As mentioned in \Sec{sec:linearframe}, in the linear \kahler frame there is a nonzero value of $\langle b_0 \rangle$. Here we calculate the values of all the SUGRA auxiliary fields in this frame. We repeat the \kahler potential and superpotential in this frame for convenience:
\be
\mb{K} = \XNL^\dagger \XNL + \alpha \BNL + \BNL^2, \qquad \mb{W} = \Mpl^2 |m_{3/2}| + f \XNL.
\ee
Following \Ref{wess1992supersymmetry}, we begin with the equations of motion for $F$ and $M$ (the scalar auxiliary field), dropping terms involving $\BNL$ for now:
\begin{align}
 &F=-f^*\left ( 1 + \frac{2}{3}|x|^2\right) - x \Mpl^2 |m_{3/2}|, \\
&M=\frac{9 \Mpl^2 |m_{3/2}| +3F^* x + 9fx}{|x|^2 -3\Mpl^2},
\end{align}
where $x$ is the $\Theta=0$ component of $\XNL$ and will eventually be replaced by $x= \xi^2/2F$ as in \Eq{eq:xnl-def}. Taking vevs of both sides, we find
\begin{align}
&\langle F \rangle = -f^*,  \label{SUGRA-vev-F}\\
 &\langle M \rangle =-3 |m_{3/2}|, \label{SUGRA-vev-M}
\end{align}
under the assumption that $\langle x \rangle = 0$; that is, there is no goldstino condensate. Note that because every piece of the $F$-term of $\BNL$ contains two or more goldstinos, restoring $\BNL$ would not change \Eqs{SUGRA-vev-F}{SUGRA-vev-M}.

The auxiliary vector equation of motion can be read from Eq. (21.15) of \Ref{wess1992supersymmetry}:
\be
b_\mu = -\frac{3i}{2}\left(\Omega_A \partial_\mu A - \Omega_{A*}\partial_\mu A^* + \Omega_x \partial_\mu x - \Omega_{x*}\partial_\mu x^* \right)\Omega^{-1} + {\rm fermions},
\ee
where $\Omega = K - 3\Mpl^2$ is evaluated on the lowest components $A$ and $x$ of the chiral superfields $\ANL$ and $\XNL$, respectively, and $\Omega_A = \partial \Omega/\partial A$, $\Omega_x = \partial \Omega/\partial x$. Taking vevs of both sides, we see that only the linear terms in the \kahler potential contribute to the term in parentheses. Once again assuming no goldstino condensates, we have
\be
\langle b_\mu \rangle = \frac{i\alpha}{2\Mpl^2}\langle \partial_\mu \varphi \rangle,
\ee
as advertised in \Eq{eq:b0advert}.

It is interesting to compare our results to the class of theories studied in \Ref{Baumann:2011nk}. There, the authors claim the EFT of inflation implies $\langle M \rangle = 0$ exactly, because the action has an $R$-symmetry, but no $R$-symmetry breaking parameters are available. We see from \Eq{SUGRA-vev-M} that the required $R$-symmetry breaking parameter is $|m_{3/2}|$. Of course, \Ref{Baumann:2011nk} considers the $|m_{3/2}| \ll H$ limit, so $\langle M \rangle \simeq 0$ is still consistent. In principle, including a nonzero $|m_{3/2}|$ is important for phenomenological reasons: SUSY remains broken after inflation ends, and to obtain a nearly-vanishing vacuum energy today, the positive vacuum energy from SUSY breaking must be offset by a fine-tuning of the negative vacuum energy from $\langle \mb{W} \rangle = \Mpl^2 |m_{3/2}|$.

\section{Fermions and time-dependent mass phases}
\label{app:alphadispersion}

As discussed in \Sec{sec:alpharole}, we consider here the effect of time-dependent phases for fermion masses. We first work out a toy example of a single Majorana fermion in \Sec{app:alphadispersion-majorana}, followed by the analysis for a gravitino in flat space in \Sec{app:alphadispersion-RS}.
\subsection{Majorana fermion} \label{app:alphadispersion-majorana}
 Consider a single Majorana fermion with a time-dependent mass phase,
\be
\mathcal{L}_{1/2} = i \bar{\xi} \sigmabar^\mu \partial_\mu \xi - \frac{m}{2}(e^{-2 i \tilde{\alpha} t}\xi \xi + e^{+2 i \tilde{\alpha} t}\bar{\xi} \bar{\xi}), \label{majorana-action-phase-mass}
\ee
where $m$ is a real parameter.\footnote{We denote the phase $\tilde{\alpha}$ here to differentiate from the $\alpha$ used in the text, which has a specific interpretation in terms of the SUGRA \kahler potential.} Under a field redefinition $\xi \to e^{i \tilde{\alpha} t} \xi$, the Lagrangian goes to
\be
\mathcal{L}_{1/2} \to \mathcal{L}'_{1/2} = i \bar{\xi} \sigmabar^\mu \partial_\mu \xi - \tilde{\alpha} \bar{\xi} \sigmabar^0 \xi -\frac{m}{2}(\xi \xi +  \bar{\xi} \bar{\xi}).
\ee
The second term is a Lorentz violating fermion bilinear, similar to the one considered in \Sec{sec:SUGRA}. If $\xi$ were a Dirac fermion, this term would be an ordinary chemical potential proportional to $\tilde{\alpha}$.  From \Eq{majorana-action-phase-mass} and our field redefinition, it should be clear that such a bilinear can be eliminated if the fermion mass is vanishing.

However, for a massive field, the phase is physical. Computing the equation of motion for $\bar{\xi}$, we can solve for $\xi$ in terms of $\bar{\xi}$. Substituting that expression into the equation of motion for $\xi$, we find
\be
\partial^2 \bar{\xi} - m^2 \bar{\xi} -\tilde{\alpha}^2\sigmabar^0 \sigma^0 \bar{\xi} +i \tilde{\alpha} (\sigmabar^0 \sigma^\mu - \sigmabar^\mu \sigma^0) \partial_\mu \bar{\xi} =0.
\ee
 Expanding in modes $\bar{\xi}(t,\vec{x}) \sim \int d\omega \, d^3 k \, e^{i\omega t -i \vec{k}\cdot \vec{x}} \bar{\xi}(\omega, \vec{k})$, we find that they must satisfy the dispersion relation
 \be
\omega^2 =  (|\vec{k}| \pm \tilde{\alpha})^2 +m^2.
\label{eq:MajDisp}
 \ee
The time-dependent phase acts as a uniform shift in the magnitude of the wavevector for each mode, with left- and right-handed eigenspinors shifting in opposite directions.  Amusingly, this allows for a negative group velocity for one of the modes at small $|\vec{k}|$. Indeed, the opposite shifts for left- and right-handed spinors is similar to the effect expected from a chemical potential, which shifts the energy levels of particles and antiparticles in opposite directions.

If we express the eigenspinors as functions of $\vec{k}$, then they are independent of $\tilde{\alpha}$, in contrast to the frame of \Eq{majorana-action-phase-mass}, where the eigenspinors would have explicit time dependence.  We emphasize that we are considering $\vec{k}$ as the independent variable when constructing the eigenspinors, with $\omega$ related to $\vec{k}$ by the dispersion relation in \Eq{eq:MajDisp}.

\subsection{Gravitino}\label{app:alphadispersion-RS}

We now consider the case of a gravitino, with the same time-dependent mass phase.  For simplicity, we work in flat space:
\be
\mathcal{L}_{3/2} = \epsilon^{\mu \nu \rho \kappa}\psibar_\mu \sigmabar_\nu \partial_\rho \psi_\kappa - m\left (e^{-2i\tilde{\alpha} t}\psi_\mu \sigma^{\mu \nu} \psi_\nu + e^{+2i\tilde{\alpha} t} \psibar_\mu \sigmabar^{\mu \nu} \psibar_\nu\right). 
\ee
Similar to the Majorana fermion, a field redefinition $\psi_\mu \to e^{i \tilde{\alpha} t} \psi_\mu$ removes the phase at the cost of a Lorentz-violating fermion bilinear.
\be
\mathcal{L}_{3/2} \to \mathcal{L}'_{3/2} = \epsilon^{\mu \nu \rho \kappa}\psibar_\mu \sigmabar_\nu \partial_\rho \psi_\kappa + i\tilde{\alpha} \epsilon^{0 \mu \nu \kappa}\psibar_\mu \sigmabar_\nu  \psi_\kappa - m\left (\psi_\mu \sigma^{\mu \nu} \psi_\nu + \psibar_\mu \sigmabar^{\mu \nu} \psibar_\nu \right).
\label{eq:32redef}
\ee
As in \Sec{sec:gravdispersion}, it is simplest to combine the two Weyl spinors $\psi_\mu$ and $\psibar_\mu$ into a 4-component spinor $\Psi^\mu$.

The modified Rarita-Schwinger equation (\ref{eq:32redef}) changes the two algebraic constraints on $\Psi^\mu_{\vec{k}}$ in a rather complicated way. Rather than solve the constraints to derive the equations of motion algebraically, we simply Fourier transform and use the equations of motion implied by \Eq{eq:32redef} to find the eigenvalues and eigenspinors of the mode matrix. Isolating the four physical degrees of freedom, we find four dispersions:
\begin{align}
\omega^2 &= (|\vec{k}| \pm \tilde{\alpha})^2 + m^2, \\
\omega^2 &= |\vec{k}| ^2 \left(1 +  \frac{16  \tilde{\alpha} ^2}{ 9 m^2} \right) \pm \frac{2}{3} \tilde{\alpha}  |\vec{k} | + m^2+ \tilde{\alpha} ^2.
\end{align}
The first pair roughly correspond to the $\psi_{3/2}$ spin polarizations, and the shift in $|\vec{k}|$ is identical to the dispersion for the Majorana fermion in \App{app:alphadispersion-majorana}. The second pair roughly correspond to the $\psi_{1/2}$ spin polarizations.

This second dispersion relation is rather unusual, and these modes eventually have a superluminal group velocity for large enough momentum. However, it is worth remembering that the flat-space case worked out in this appendix is just a simplified toy example. In an expanding universe, this relation will be modified, since these modes already have a reduced speed of sound given by \Eq{gravitino-small-cs}, and the parameter $\alpha$ of our EFT is both slow-roll and Planck suppressed.  In particular, $\tilde{\alpha} = \alpha \sqrt{\epsilon} H/ 2 \sqrt{2} \Mpl$, so if one naively combines the effect of $\alpha$ with the reduced sound speed, one only finds superluminal propagation if $\alpha \gtrsim \Mpl \gg \Lambda$, which violates the EFT power counting.

\bibliographystyle{JHEP}
\bibliography{SUSYInflationBib}{}

\providecommand{\href}[2]{#2}\begingroup\raggedright\begin{thebibliography}{10}

\bibitem{Guth:1980zm}
A.~H. Guth, {\it {The Inflationary Universe: A Possible Solution to the Horizon
  and Flatness Problems}},  {\em Phys.Rev.} {\bf D23} (1981) 347--356.

\bibitem{Linde:1981mu}
A.~D. Linde, {\it {A New Inflationary Universe Scenario: A Possible Solution of
  the Horizon, Flatness, Homogeneity, Isotropy and Primordial Monopole
  Problems}},  {\em Phys.Lett.} {\bf B108} (1982) 389--393.

\bibitem{Lyth:1998xn}
D.~H. Lyth and A.~Riotto, {\it {Particle physics models of inflation and the
  cosmological density perturbation}},  {\em Phys.Rept.} {\bf 314} (1999)
  1--146, [\href{http://arxiv.org/abs/hep-ph/9807278}{{\tt hep-ph/9807278}}].

\bibitem{2011CQGra..28j3001Y}
M.~{Yamaguchi}, {\it {Supergravity-based inflation models: a review}},  {\em
  Classical and Quantum Gravity} {\bf 28} (May, 2011) 103001,
  [\href{http://arxiv.org/abs/1101.2488}{{\tt arXiv:1101.2488}}].

\bibitem{Kachru:2003aw}
S.~Kachru, R.~Kallosh, A.~D. Linde, and S.~P. Trivedi, {\it {De Sitter vacua in
  string theory}},  {\em Phys.Rev.} {\bf D68} (2003) 046005,
  [\href{http://arxiv.org/abs/hep-th/0301240}{{\tt hep-th/0301240}}].

\bibitem{Balasubramanian:2005zx}
V.~Balasubramanian, P.~Berglund, J.~P. Conlon, and F.~Quevedo, {\it
  {Systematics of moduli stabilisation in Calabi-Yau flux compactifications}},
  {\em JHEP} {\bf 0503} (2005) 007,
  [\href{http://arxiv.org/abs/hep-th/0502058}{{\tt hep-th/0502058}}].

\bibitem{Baumann:2014nda}
D.~Baumann and L.~McAllister, {\it {Inflation and String Theory}},
  \href{http://arxiv.org/abs/1404.2601}{{\tt arXiv:1404.2601}}.

\bibitem{Cheung:2007st}
C.~Cheung, P.~Creminelli, A.~L. Fitzpatrick, J.~Kaplan, and L.~Senatore, {\it
  {The Effective Field Theory of Inflation}},  {\em JHEP} {\bf 0803} (2008)
  014, [\href{http://arxiv.org/abs/0709.0293}{{\tt arXiv:0709.0293}}].

\bibitem{Senatore:2010wk}
L.~Senatore and M.~Zaldarriaga, {\it {The Effective Field Theory of Multifield
  Inflation}},  {\em JHEP} {\bf 1204} (2012) 024,
  [\href{http://arxiv.org/abs/1009.2093}{{\tt arXiv:1009.2093}}].

\bibitem{Baumann:2011nk}
D.~Baumann and D.~Green, {\it {Signatures of Supersymmetry from the Early
  Universe}},  {\em Phys.Rev.} {\bf D85} (2012) 103520,
  [\href{http://arxiv.org/abs/1109.0292}{{\tt arXiv:1109.0292}}].

\bibitem{Baumann:2011nm}
D.~Baumann and D.~Green, {\it {Supergravity for Effective Theories}},  {\em
  JHEP} {\bf 1203} (2012) 001, [\href{http://arxiv.org/abs/1109.0293}{{\tt
  arXiv:1109.0293}}].

\bibitem{Komargodski:2009rz}
Z.~Komargodski and N.~Seiberg, {\it {From Linear SUSY to Constrained
  Superfields}},  {\em JHEP} {\bf 0909} (2009) 066,
  [\href{http://arxiv.org/abs/0907.2441}{{\tt arXiv:0907.2441}}].

\bibitem{Cheung:2010mc}
C.~Cheung, Y.~Nomura, and J.~Thaler, {\it {Goldstini}},  {\em JHEP} {\bf 1003}
  (2010) 073, [\href{http://arxiv.org/abs/1002.1967}{{\tt arXiv:1002.1967}}].

\bibitem{AlvarezGaume:2010rt}
L.~Alvarez-Gaume, C.~Gomez, and R.~Jimenez, {\it {Minimal Inflation}},  {\em
  Phys.Lett.} {\bf B690} (2010) 68--72,
  [\href{http://arxiv.org/abs/1001.0010}{{\tt arXiv:1001.0010}}].

\bibitem{AlvarezGaume:2011xv}
L.~Alvarez-Gaume, C.~Gomez, and R.~Jimenez, {\it {A Minimal Inflation
  Scenario}},  {\em JCAP} {\bf 1103} (2011) 027,
  [\href{http://arxiv.org/abs/1101.4948}{{\tt arXiv:1101.4948}}].

\bibitem{AlvarezGaume:2011db}
L.~Alvarez-Gaume, C.~Gomez, and R.~Jimenez, {\it {Phenomenology of the minimal
  inflation scenario: inflationary trajectories and particle production}},
  {\em JCAP} {\bf 1203} (2012) 017, [\href{http://arxiv.org/abs/1110.3984}{{\tt
  arXiv:1110.3984}}].

\bibitem{Ellis:2013zsa}
J.~Ellis and N.~E. Mavromatos, {\it {Inflation induced by gravitino
  condensation in supergravity}},  {\em Phys.Rev.} {\bf D88} (2013), no.~8
  085029, [\href{http://arxiv.org/abs/1308.1906}{{\tt arXiv:1308.1906}}].

\bibitem{Antoniadis:2014oya}
I.~Antoniadis, E.~Dudas, S.~Ferrara, and A.~Sagnotti, {\it {The
  Volkov-Akulov-Starobinsky supergravity}},  {\em Phys.Lett.} {\bf B733} (2014)
  32--35, [\href{http://arxiv.org/abs/1403.3269}{{\tt arXiv:1403.3269}}].

\bibitem{Ferrara:2014kva}
S.~Ferrara, R.~Kallosh, and A.~Linde, {\it {Cosmology with Nilpotent
  Superfields}},  {\em JHEP} {\bf 1410} (2014) 143,
  [\href{http://arxiv.org/abs/1408.4096}{{\tt arXiv:1408.4096}}].

\bibitem{Kallosh:2014via}
R.~Kallosh and A.~Linde, {\it {Inflation and Uplifting with Nilpotent
  Superfields}},  {\em JCAP} {\bf 1501} (2015), no.~01 025,
  [\href{http://arxiv.org/abs/1408.5950}{{\tt arXiv:1408.5950}}].

\bibitem{DallAgata:2014oka}
G.~Dall'Agata and F.~Zwirner, {\it {On sgoldstino-less supergravity models of
  inflation}},  {\em JHEP} {\bf 1412} (2014) 172,
  [\href{http://arxiv.org/abs/1411.2605}{{\tt arXiv:1411.2605}}].

\bibitem{Kallosh:2014hxa}
R.~Kallosh, A.~Linde, and M.~Scalisi, {\it {Inflation, de Sitter Landscape and
  Super-Higgs effect}},  \href{http://arxiv.org/abs/1411.5671}{{\tt
  arXiv:1411.5671}}.

\bibitem{Mavromatos:2014yaa}
N.~E. Mavromatos, {\it {Gravitino Condensates in the Early Universe and
  Inflation}},  \href{http://arxiv.org/abs/1412.6437}{{\tt arXiv:1412.6437}}.

\bibitem{Arkani-Hamed:2015bza}
N.~Arkani-Hamed and J.~Maldacena, {\it {Cosmological Collider Physics}},
  \href{http://arxiv.org/abs/1503.08043}{{\tt arXiv:1503.08043}}.

\bibitem{Ade:2014xna}
{\bf BICEP2 Collaboration} Collaboration, P.~Ade et~al., {\it {Detection of
  $B$-Mode Polarization at Degree Angular Scales by BICEP2}},  {\em
  Phys.Rev.Lett.} {\bf 112} (2014), no.~24 241101,
  [\href{http://arxiv.org/abs/1403.3985}{{\tt arXiv:1403.3985}}].

\bibitem{Ade:2015tva}
{\bf BICEP2 Collaboration, Planck Collaboration} Collaboration, P.~Ade et~al.,
  {\it {A Joint Analysis of BICEP2/Keck Array and Planck Data}},  {\em
  Phys.Rev.Lett.} (2015) [\href{http://arxiv.org/abs/1502.00612}{{\tt
  arXiv:1502.00612}}].

\bibitem{Senatore:2009cf}
L.~Senatore and M.~Zaldarriaga, {\it {On Loops in Inflation}},  {\em JHEP} {\bf
  1012} (2010) 008, [\href{http://arxiv.org/abs/0912.2734}{{\tt
  arXiv:0912.2734}}].

\bibitem{Chen:2009zp}
X.~Chen and Y.~Wang, {\it {Quasi-Single Field Inflation and
  Non-Gaussianities}},  {\em JCAP} {\bf 1004} (2010) 027,
  [\href{http://arxiv.org/abs/0911.3380}{{\tt arXiv:0911.3380}}].

\bibitem{2012JCAP...08..033S}
E.~{Sefusatti}, J.~R. {Fergusson}, X.~{Chen}, and E.~P.~S. {Shellard}, {\it
  {Effects and detectability of quasi-single field inflation in the large-scale
  structure and cosmic microwave background}},  {\em JCAP} {\bf 8} (Aug., 2012)
  33, [\href{http://arxiv.org/abs/1204.6318}{{\tt arXiv:1204.6318}}].

\bibitem{2012JCAP...08..019N}
J.~{Nore{\~n}a}, L.~{Verde}, G.~{Barenboim}, and C.~{Bosch}, {\it {Prospects
  for constraining the shape of non-Gaussianity with the scale-dependent
  bias}},  {\em JCAP} {\bf 8} (Aug., 2012) 19,
  [\href{http://arxiv.org/abs/1204.6324}{{\tt arXiv:1204.6324}}].

\bibitem{Assassi:2013gxa}
V.~Assassi, D.~Baumann, D.~Green, and L.~McAllister, {\it {Planck-Suppressed
  Operators}},  {\em JCAP} {\bf 1401} (2014), no.~01 033,
  [\href{http://arxiv.org/abs/1304.5226}{{\tt arXiv:1304.5226}}].

\bibitem{Craig:2014rta}
N.~Craig and D.~Green, {\it {Testing Split Supersymmetry with Inflation}},
  {\em JHEP} {\bf 1407} (2014) 102, [\href{http://arxiv.org/abs/1403.7193}{{\tt
  arXiv:1403.7193}}].

\bibitem{Dimastrogiovanni:2015pla}
E.~Dimastrogiovanni, M.~Fasiello, and M.~Kamionkowski, {\it {Imprints of
  Massive Primordial Fields on Large-Scale Structure}},
  \href{http://arxiv.org/abs/1504.05993}{{\tt arXiv:1504.05993}}.

\bibitem{Kehagias:2015jha}
A.~Kehagias and A.~Riotto, {\it {High Energy Physics Signatures from Inflation
  and Conformal Symmetry of de Sitter}},  {\em Fortsch. Phys.} {\bf 63} (2015)
  531--542, [\href{http://arxiv.org/abs/1501.03515}{{\tt arXiv:1501.03515}}].

\bibitem{ArkaniHamed:2003uy}
N.~Arkani-Hamed, H.-C. Cheng, M.~A. Luty, and S.~Mukohyama, {\it {Ghost
  condensation and a consistent infrared modification of gravity}},  {\em JHEP}
  {\bf 0405} (2004) 074, [\href{http://arxiv.org/abs/hep-th/0312099}{{\tt
  hep-th/0312099}}].

\bibitem{ArkaniHamed:2004ar}
N.~Arkani-Hamed, H.-C. Cheng, M.~Luty, and J.~Thaler, {\it {Universal dynamics
  of spontaneous Lorentz violation and a new spin-dependent inverse-square law
  force}},  {\em JHEP} {\bf 0507} (2005) 029,
  [\href{http://arxiv.org/abs/hep-ph/0407034}{{\tt hep-ph/0407034}}].

\bibitem{ArkaniHamed:2005gu}
N.~Arkani-Hamed, H.-C. Cheng, M.~A. Luty, S.~Mukohyama, and T.~Wiseman, {\it
  {Dynamics of gravity in a Higgs phase}},  {\em JHEP} {\bf 0701} (2007) 036,
  [\href{http://arxiv.org/abs/hep-ph/0507120}{{\tt hep-ph/0507120}}].

\bibitem{Cheng:2006us}
H.-C. Cheng, M.~A. Luty, S.~Mukohyama, and J.~Thaler, {\it {Spontaneous Lorentz
  breaking at high energies}},  {\em JHEP} {\bf 0605} (2006) 076,
  [\href{http://arxiv.org/abs/hep-th/0603010}{{\tt hep-th/0603010}}].

\bibitem{Khoury:2010gb}
J.~Khoury, J.-L. Lehners, and B.~Ovrut, {\it {Supersymmetric P(X,$\phi$) and
  the Ghost Condensate}},  {\em Phys.Rev.} {\bf D83} (2011) 125031,
  [\href{http://arxiv.org/abs/1012.3748}{{\tt arXiv:1012.3748}}].

\bibitem{Koehn:2012te}
M.~Koehn, J.-L. Lehners, and B.~Ovrut, {\it {Ghost condensate in $N=1$
  supergravity}},  {\em Phys.Rev.} {\bf D87} (2013), no.~6 065022,
  [\href{http://arxiv.org/abs/1212.2185}{{\tt arXiv:1212.2185}}].

\bibitem{Shapere:2012nq}
A.~Shapere and F.~Wilczek, {\it {Classical Time Crystals}},  {\em
  Phys.Rev.Lett.} {\bf 109} (2012) 160402,
  [\href{http://arxiv.org/abs/1202.2537}{{\tt arXiv:1202.2537}}].

\bibitem{Wilczek:2012jt}
F.~Wilczek, {\it {Quantum Time Crystals}},  {\em Phys.Rev.Lett.} {\bf 109}
  (2012) 160401, [\href{http://arxiv.org/abs/1202.2539}{{\tt
  arXiv:1202.2539}}].

\bibitem{PhysRevLett.111.070402}
P.~Bruno, {\it Impossibility of spontaneously rotating time crystals: A no-go
  theorem},  {\em Phys. Rev. Lett.} {\bf 111} (Aug, 2013) 070402.

\bibitem{Benakli:2013ava}
K.~Benakli, Y.~Oz, and G.~Policastro, {\it {The Super-Higgs Mechanism in
  Fluids}},  {\em JHEP} {\bf 1402} (2014) 015,
  [\href{http://arxiv.org/abs/1310.5002}{{\tt arXiv:1310.5002}}].

\bibitem{Benakli:2014bpa}
K.~Benakli, L.~Darm{\'e}, and Y.~Oz, {\it {The Slow Gravitino}},
  \href{http://arxiv.org/abs/1407.8321}{{\tt arXiv:1407.8321}}.

\bibitem{Giudice:1999yt}
G.~Giudice, I.~Tkachev, and A.~Riotto, {\it {Nonthermal production of dangerous
  relics in the early universe}},  {\em JHEP} {\bf 9908} (1999) 009,
  [\href{http://arxiv.org/abs/hep-ph/9907510}{{\tt hep-ph/9907510}}].

\bibitem{wess1992supersymmetry}
J.~Wess and J.~Bagger, {\em Supersymmetry and Supergravity}.
\newblock Princeton University Press, 1992.

\bibitem{Senatore:2009gt}
L.~Senatore, K.~M. Smith, and M.~Zaldarriaga, {\it {Non-Gaussianities in Single
  Field Inflation and their Optimal Limits from the WMAP 5-year Data}},  {\em
  JCAP} {\bf 1001} (2010) 028, [\href{http://arxiv.org/abs/0905.3746}{{\tt
  arXiv:0905.3746}}].

\bibitem{Sasaki:2012ka}
S.~Sasaki, M.~Yamaguchi, and D.~Yokoyama, {\it {Supersymmetric DBI inflation}},
   {\em Phys.Lett.} {\bf B718} (2012) 1--4,
  [\href{http://arxiv.org/abs/1205.1353}{{\tt arXiv:1205.1353}}].

\bibitem{Gwyn:2014wna}
R.~Gwyn and J.-L. Lehners, {\it {Non-Canonical Inflation in Supergravity}},
  {\em JHEP} {\bf 1405} (2014) 050, [\href{http://arxiv.org/abs/1402.5120}{{\tt
  arXiv:1402.5120}}].

\bibitem{DEramo:2013mya}
F.~D'Eramo, J.~Thaler, and Z.~Thomas, {\it {Anomaly Mediation from Unbroken
  Supergravity}},  {\em JHEP} {\bf 1309} (2013) 125,
  [\href{http://arxiv.org/abs/1307.3251}{{\tt arXiv:1307.3251}}].

\bibitem{Bertolini:2013via}
D.~Bertolini, J.~Thaler, and Z.~Thomas, {\it {TASI 2012: Super-Tricks for
  Superspace}},  \href{http://arxiv.org/abs/1302.6229}{{\tt arXiv:1302.6229}}.

\bibitem{Kratzert:2003cr}
K.~Kratzert, {\it {Finite temperature supersymmetry: The Wess-Zumino model}},
  {\em Annals Phys.} {\bf 308} (2003) 285--310,
  [\href{http://arxiv.org/abs/hep-th/0303260}{{\tt hep-th/0303260}}].

\bibitem{Cheung:2011jp}
C.~Cheung, F.~D'Eramo, and J.~Thaler, {\it {Supergravity Computations without
  Gravity Complications}},  {\em Phys.Rev.} {\bf D84} (2011) 085012,
  [\href{http://arxiv.org/abs/1104.2598}{{\tt arXiv:1104.2598}}].

\bibitem{Cohen:1987vi}
A.~G. Cohen and D.~B. Kaplan, {\it {Thermodynamic Generation of the Baryon
  Asymmetry}},  {\em Phys.Lett.} {\bf B199} (1987) 251.

\bibitem{Cohen:1988kt}
A.~G. Cohen and D.~B. Kaplan, {\it {SPONTANEOUS BARYOGENESIS}},  {\em
  Nucl.Phys.} {\bf B308} (1988) 913.

\bibitem{Hook:2014mla}
A.~Hook, {\it {Baryogenesis from Hawking Radiation}},  {\em Phys.Rev.} {\bf
  D90} (2014), no.~8 083535, [\href{http://arxiv.org/abs/1404.0113}{{\tt
  arXiv:1404.0113}}].

\bibitem{DAgnolo:2015pha}
R.~T. D'Agnolo and A.~Hook, {\it {Selfish Dark Matter}},
  \href{http://arxiv.org/abs/1504.00361}{{\tt arXiv:1504.00361}}.

\bibitem{Maroto:1999ch}
A.~L. Maroto and A.~Mazumdar, {\it {Production of spin 3/2 particles from
  vacuum fluctuations}},  {\em Phys. Rev. Lett.} {\bf 84} (2000) 1655--1658,
  [\href{http://arxiv.org/abs/hep-ph/9904206}{{\tt hep-ph/9904206}}].

\bibitem{Giudice:1999am}
G.~Giudice, A.~Riotto, and I.~Tkachev, {\it {Thermal and nonthermal production
  of gravitinos in the early universe}},  {\em JHEP} {\bf 9911} (1999) 036,
  [\href{http://arxiv.org/abs/hep-ph/9911302}{{\tt hep-ph/9911302}}].

\bibitem{Kallosh:1999jj}
R.~Kallosh, L.~Kofman, A.~D. Linde, and A.~Van~Proeyen, {\it {Gravitino
  production after inflation}},  {\em Phys.Rev.} {\bf D61} (2000) 103503,
  [\href{http://arxiv.org/abs/hep-th/9907124}{{\tt hep-th/9907124}}].

\bibitem{Kallosh:2000ve}
R.~Kallosh, L.~Kofman, A.~D. Linde, and A.~Van~Proeyen, {\it {Superconformal
  symmetry, supergravity and cosmology}},  {\em Class.Quant.Grav.} {\bf 17}
  (2000) 4269--4338, [\href{http://arxiv.org/abs/hep-th/0006179}{{\tt
  hep-th/0006179}}].

\bibitem{Weinberg:2005vy}
S.~Weinberg, {\it {Quantum contributions to cosmological correlations}},  {\em
  Phys.Rev.} {\bf D72} (2005) 043514,
  [\href{http://arxiv.org/abs/hep-th/0506236}{{\tt hep-th/0506236}}].

\bibitem{Senatore:2012nq}
L.~Senatore and M.~Zaldarriaga, {\it {On Loops in Inflation II: IR Effects in
  Single Clock Inflation}},  {\em JHEP} {\bf 01} (2013) 109,
  [\href{http://arxiv.org/abs/1203.6354}{{\tt arXiv:1203.6354}}].

\bibitem{Pimentel:2012tw}
G.~L. Pimentel, L.~Senatore, and M.~Zaldarriaga, {\it {On Loops in Inflation
  III: Time Independence of zeta in Single Clock Inflation}},  {\em JHEP} {\bf
  07} (2012) 166, [\href{http://arxiv.org/abs/1203.6651}{{\tt
  arXiv:1203.6651}}].

\bibitem{Senatore:2012ya}
L.~Senatore and M.~Zaldarriaga, {\it {The constancy of $\zeta$ in single-clock
  Inflation at all loops}},  {\em JHEP} {\bf 09} (2013) 148,
  [\href{http://arxiv.org/abs/1210.6048}{{\tt arXiv:1210.6048}}].

\bibitem{Assassi:2012et}
V.~Assassi, D.~Baumann, and D.~Green, {\it {Symmetries and Loops in
  Inflation}},  {\em JHEP} {\bf 1302} (2013) 151,
  [\href{http://arxiv.org/abs/1210.7792}{{\tt arXiv:1210.7792}}].

\bibitem{Green:2013rd}
D.~Green, M.~Lewandowski, L.~Senatore, E.~Silverstein, and M.~Zaldarriaga, {\it
  {Anomalous Dimensions and Non-Gaussianity}},  {\em JHEP} {\bf 1310} (2013)
  171, [\href{http://arxiv.org/abs/1301.2630}{{\tt arXiv:1301.2630}}].

\bibitem{Anninos:2014lwa}
D.~Anninos, T.~Anous, D.~Z. Freedman, and G.~Konstantinidis, {\it {Late-time
  Structure of the Bunch-Davies De Sitter Wavefunction}},
  \href{http://arxiv.org/abs/1406.5490}{{\tt arXiv:1406.5490}}.

\bibitem{Marolf:2010zp}
D.~Marolf and I.~A. Morrison, {\it {The IR stability of de Sitter: Loop
  corrections to scalar propagators}},  {\em Phys.Rev.} {\bf D82} (2010)
  105032, [\href{http://arxiv.org/abs/1006.0035}{{\tt arXiv:1006.0035}}].

\bibitem{Krotov:2010ma}
D.~Krotov and A.~M. Polyakov, {\it {Infrared Sensitivity of Unstable Vacua}},
  {\em Nucl.Phys.} {\bf B849} (2011) 410--432,
  [\href{http://arxiv.org/abs/1012.2107}{{\tt arXiv:1012.2107}}].

\bibitem{Maldacena:2002vr}
J.~M. Maldacena, {\it {Non-Gaussian features of primordial fluctuations in
  single field inflationary models}},  {\em JHEP} {\bf 0305} (2003) 013,
  [\href{http://arxiv.org/abs/astro-ph/0210603}{{\tt astro-ph/0210603}}].

\bibitem{Ghosh:2014kba}
A.~Ghosh, N.~Kundu, S.~Raju, and S.~P. Trivedi, {\it {Conformal Invariance and
  the Four Point Scalar Correlator in Slow-Roll Inflation}},  {\em JHEP} {\bf
  1407} (2014) 011, [\href{http://arxiv.org/abs/1401.1426}{{\tt
  arXiv:1401.1426}}].

\bibitem{Kundu:2014gxa}
N.~Kundu, A.~Shukla, and S.~P. Trivedi, {\it {Constraints from Conformal
  Symmetry on the Three Point Scalar Correlator in Inflation}},
  \href{http://arxiv.org/abs/1410.2606}{{\tt arXiv:1410.2606}}.

\bibitem{Nilles:2001ry}
H.~P. Nilles, M.~Peloso, and L.~Sorbo, {\it {Nonthermal production of
  gravitinos and inflatinos}},  {\em Phys.Rev.Lett.} {\bf 87} (2001) 051302,
  [\href{http://arxiv.org/abs/hep-ph/0102264}{{\tt hep-ph/0102264}}].

\end{thebibliography}\endgroup

\end{document}